\documentstyle[12pt]{article}
\title{The stability of a zonally averaged\\
thermohaline circulation model}
\author{G.A. Schmidt \   and \  L.A. Mysak \\ \\
Centre for Climate and Global Change Research\\
and \\
Department of Atmospheric and Oceanic Sciences\\
McGill University \\
805 Sherbrooke St. W, Montreal, QC, H3A 2K6, Canada \\ }
\date{ Revised for Tellus \\ Mar 1995 }
\newcommand{\p}{\partial}
\newcommand{\head}[1]{\vskip 3ex \noindent{\bf #1} \vskip 3ex}
\newcommand{\subhead}[1]{\vskip 3ex \noindent{\it #1} \vskip 3ex}
\newcommand{\tbl}[4]{\renewcommand{\baselinestretch}{1.0}\small
\begin{figure}[t] \begin{center} \begin{quote}{\bf Table #3:}
#4\end{quote} \vskip 5ex \begin{tabular}{#1}\hline #2 \hline
\end{tabular} \end{center}
\end{figure}\renewcommand{\baselinestretch}{1.0}\normalsize}
\newcommand{\fig}[5]{
\renewcommand{\baselinestretch}{1.0}\small
\begin{figure}[h]\begin{picture}(#1)\put(#2){\includegraphics{fig#3.ps}}\end{picture}\begin{quote}{\bf
Figure #3}: #5
\end{quote}\end{figure}\renewcommand{\baselinestretch}{1.0}\normalsize}

\renewcommand{\baselinestretch}{1.0}
\begin{document}

\maketitle
\newpage

\centerline{ABSTRACT}
\vskip 3ex

A combination of analytical and numerical techniques are used to
efficiently determine  the
qualitative and quantitative behaviour of a one-basin zonally
averaged thermohaline circulation ocean model under various
forcing regimes and over a large region of parameter space.
In contrast to earlier studies which use time stepping to find the
steady solutions, the steady state equations are first solved directly
to obtain the multiple equilibria under identical mixed boundary conditions.
This approach is based on the differentiability of the governing equations
and especially the convection scheme. A linear stability analysis is then
performed, in which the normal modes and corresponding eigenvalues
are found for the various equilibrium states. Resonant periodic solutions
superimposed on these states are predicted for
various types of forcing.

The results are used to gain insight into the solutions obtained by
Mysak, Stocker and Huang in a previous numerical study in which the
eddy diffusivities were varied in a randomly forced one-basin zonally
averaged model. It is shown that  the two-cell symmetric
circulation is  generally unstable to anti-symmetric perturbations in the
temperature or salinity (as expected), and that both one-cell
(inter-hemispheric) circulation patterns are generally stable.
Resonant stable oscillations with
century scale periods are predicted with structures that
compare favorably with those found in the previous study. In cases with large
horizontal diffusivities, the two-cell pattern is also stable, which
parallels cases in the previous study where large vacillations were
seen between the three stable steady states. Further, in cases with
large horizontal {\it and} large vertical diffusivities, no one-cell
pattern can be realised and the only asymptotic behaviour
found is the two-cell pattern.

An experiment is also performed to examine the effect of varying the
restoring time constants in the relaxation boundary conditions used at
the surface for both salinity and temperature.

\newpage

\head{1. Introduction}

Multiple equilibria  of the thermohaline ocean circulation (THC) under mixed
boundary conditions (restoring condition for temperature, a flux condition for
salinity) have been found in many recent theoretical studies. In
examples ranging
from simple Stommel-type box models (Stommel, 1961; Cessi, 1994)
to full general circulation models (GCMs)
(Manabe and Stouffer, 1988; Weaver and Sarachik, 1991), various steady
states of the circulation can
exist under identical forcing. In the box models, both temperature and
salinity dominant states are found, with one circulation being the reverse
of the other. In full GCMs, depending on the geometry used, both
northern sinking and southern sinking states are among those seen (see
Weaver and Hughes (1992) for a review).

A question that arises is how relevant are these models (and the
various THC states) for the
actual thermohaline circulation? Also, how robust are the multiple
equilibria, and how much do they depend upon the various
simplifications and parameterisations used in the models? If the models
are relevant, then the question of the stability of the equilibria
becomes crucial.

The past approach to answering these questions has either been
analytic (from a non-linear
dynamics standpoint) or purely numeric. The analytic approach can
provide a complete description of the behaviour in only the relatively
simple models (see for instance Maas (1994)). The numeric
approach (Bryan, 1986) provides much less complete
information, but it can deal with much more complex cases. One barrier to
providing a more complete determination of the behaviour of the
complex models is the
exhausting (computational) effort needed to explore the full model
parameter space which is frequently  6- or 7-dimensional. Another factor
to be considered is the sensitivity of the solutions to the initial
conditions.

This paper demonstrates a {\it methodology} that can be used to
analyse the stability of the more
complicated numerical models by invoking analytical techniques more
normally applied to systems with many fewer degrees of freedom.
Many models can be accurately characterised by the number
and nature of the steady states of the model. Hence by using a
methodology which concentrates
on these states, it is hoped that the effect of parameter changes and
different initial conditions can be more clearly seen.

To illustrate the advantages of this methodology, the stability of the
Wright and Stocker (1991) zonally averaged THC
model is investigated. This model has been successfully employed in
various ocean climate (Stocker and Wright, 1991; Wright and Stocker,
1992) and paleoclimatic (Stocker{\it et al}, 1992) studies. When under
random forcing (Mysak {\it et al}, 1993)
(hereafter MSH), the one-basin model exhibits century scale fluctuations
superimposed on a basic one-cell circulation pattern (with either northern
sinking or southern sinking).  For large horizontal eddy
diffusivities a two-cell symmetric circulation can persist.

In contrast to the numerical approach used in the past in which the
model is time stepped to equilibrium, here the various steady
solutions are found {\it directly}.
The steady state equations are first solved under restoring
conditions and a salinity flux diagnosed as before. Then the steady
equations are solved under the new flux condition for salinity. The
methodology which is fully explained in \S2 centrally depends on the
differentiability of the governing equations and specifically the
convection scheme. It should be added that since
this is a highly non-linear problem, there is always some
uncertainty as to whether all the steady states have been found.
Isolated steady states often exist which
cannot be found simply by tracing the path of the known steady states
through the parameter space.  In this case there is possibly another
steady state that has a collapsed thermohaline circulation (i.e., one
with no deep water formation). However,
this is difficult to stimulate and hence there is doubt as to
whether it exists in the one-basin model used here (though it has been
seen in a two-basin model (Stocker and Wright, 1991)).

Once a steady state has been found, a linear stability analysis is
performed. This will determine whether the steady state found is
stable or not, and if not, what are the fastest growing modes. Also,
information is obtained about the nature of possible oscillations (growing,
neutrally stable or damped) and the frequencies associated with them.
These modes and the associated frequencies provide all the
information needed to describe the small amplitude behaviour of the
model under various forcing regimes around a given steady state. For a
particular forcing frequency, the complete solution for the resultant
periodic solution can be written down. Certain frequencies will cause
a resonant response, and it is those frequencies that will be most
evident under stochastic (white-noise) forcing. In this way, a
qualitative picture can be
built up of the behaviour of the model which is hopefully more robust
than the quantative results gained from simply running the model to
equilibrium in different parameter regimes and from various initial states.

As various parameters change, the stability criteria will also
change; the points in parameter space at which these changes occur
(bifurcations) are also where
new forms of asymptotic behaviour can emerge. For instance, there can be a
change from stability to instability (through a a pitchfork
bifurcation) implying new steady states. Similarly, a change from
damped to growing oscillations (through a Hopf
bifurcation) implies the existence
of a finite limit cycle, although the analysis of the resulting
periodic solutions requires more sophisticated tools than will be
presented here.

If this analysis can be used to explain the behaviour in an
already investigated problem, then it can be used to {\em predict} the
behaviour in different parameter regimes or with different boundary
conditions without the need to run the model for thousands of computer years.

As a first test, the methodology is applied to the work of MSH
mentioned above. In that paper a small-amplitude stochastic forcing
was added to the freshwater flux in each surface grid cell,
and the eddy diffusivities were varied over a wide range. In a
large region of parameter space, oscillations  with century-scale
periods were seen to be induced around various steady states. The
results in \S3
indicate that these oscillations can be accurately represented by
resonant modes of a linearised model centred on the steady states.
MSH also showed that very long (millennial) period vacillations could
be induced between steady states (including the two-cell circulation).
The results here indicate that these vacillations are only induced at
points at which the intermediate two-cell circulation is stable and
where there are three competing basins of attraction.

A second set of experiments are presented in which the
restoring times used in the temperature and salinity boundary
conditions were varied.
In the past, investigators have followed Haney (1971) who
used heat and energy flux arguments to derive a relaxation boundary
condition for temperature which used a time constant of around 50
days. His argument amounted to assuming that the atmosphere
had an infinite heat capacity (i.e.,
the surface air temperature does not change). It has been pointed out
by Schopf (1983) and by Zhang {\em et al} (1993) that in
comparison with the ocean, the atmosphere has a very small heat
capacity and that therefore a zero heat capacity atmosphere might be more
suitable. Such a model also implies that
a restoring boundary condition for temperature should be used but with
a much longer restoring time, around 600 days. Use of this formulation has been
shown to prevent the collapse of the thermohaline circulation upon the
switch to mixed boundary conditions in GCMs (the so-called polar
halocline catastrophe) (Zhang {\it et al}, 1993). This collapse is
similar to the
what happpens to the initial two-cell circulation in the one-basin zonally
averaged model, but with the crucial difference that the two-cell
pattern is not close to the observed state of the present THC.

In the spin up of ocean models (including GCMs) a restoring condition
is also used for the salinity. This has no physical basis and should be
seen merely as data assimilation in order to deduce a salinity flux
(evaporation/precipitation) at some steady state similar to today's ocean.
The restoring time used for this is usually the same as for the
temperature. However, Tziperman {\em et al} (1994)
have recently argued that such a short restoring time
constrains the salinity field too highly.
They argue that given the accuracy to which the surface salinity
concentration is known, a restoring time of at least 120 days should be
used. As above, the collapse of the thermohaline
circulation in GCMs upon a change to mixed boundary conditions, can
be avoided with such a restoring time.

The stability of the solutions for various values for both restoring
time constants are examined using a similar
procedure as followed in Experiment 1. As was found for the GCMs, a
longer restoring time for temperature does lead to a more stable
state, but the qualitative features of the solutions remain unaffected.

This paper is structured as follows. In \S 2 an outline is given of how
a local linearisation of the model can be used to solve the steady
state equations and how the behaviour of
the model near a steady state can be approximated. The results from
the first experiment, where the eddy diffusivities are varied over a
large range (following MSH), are contained in \S 3.

Section 4 describes the results from the second set of experiments
(varying the restoring time constants). In \S 5 a general discussion
of the methodology used and its utility for further studies is given.
In an appendix some of the more technical aspects
of the new convection scheme used are derived and discussed.
\head{\bf 2. The model and experimental procedure}

The model used in this study is a variant of that
developed by Wright and Stocker (1991) (hereafter WS). This
is a zonally averaged model and since the ocean basin has boundaries,
there remains a term in the equations related to the east-west
pressure gradient. To close the problem, WS assume that this can be
written as a function of the north-south pressure gradient with
constant of proportionality $\epsilon$, the pressure gradient closure
parameter (see p1716 in WS for further details). The
most significant difference in the model used here is that a new
convection scheme is used. The original WS model (in common with many
GCMs (Marotzke, 1991)) uses a highly non-linear scheme at each time
step which mixes boxes depending
on the static stability at each interface. This process can take a
number of passes through a column to obtain a statically stable state.
There are a
number of disadvantages to this process in terms of finding steady
states and determining the linear stability. For example, the mixing
is an analogue for the amount of convection that would take place in a
given time step; hence if an attempt is to be made to determine steady
states {\it without time stepping the model}, i.e. with $\Delta t=0$,
then this procedure is not helpful. The step-funtion-like behaviour at
every point in this scheme (full convection or no convection) means that it
is very non-linear and non-differentiable. As a consequence, it is
impossible to determine
analytically the effect of a small perturbation in the temperature or
salinity fields on the convective flux. A more suitable
parameterisation would be one in which the convective heat and salt
transports were independent of the time step used and differentiable
with respect to density in each box. One such scheme (derived and
discussed in Appendix A) has been used
throughout this paper. This uses a vertical diffusion coefficient
($K_v$) which varies smoothly from a small value (indicating eddy
diffusion) to a large spatially-varying one (representing convection)
as the density gradient at the interface between two boxes changes.
The scheme is differentiable at every point with respect to the
densities above and below the interface.

The convection scheme used here leads to slightly altered flows as
compared to those in WS; however,the conclusions of previous work
still hold. All the other parameters, i.e. the surface restoring time,
the surface boundary conditions, the closure parameter and the
resolution, are identical to those considered previously.

As mentioned in the introduction, the steady states of the model are
found directly
rather than by time-stepping the equations. This is a procedure
commonly followed in studies of dynamical systems and is
straightforward once the spatial parts of the equations have been
discretised (using the method outlined in Fiadiero and Veronis (1977)
and extended by Wright (1992)). The original infinite-dimensional
partial differential system is approximated by an $N$-dimensional
system, where $N=2mn$ and $m,n$ are the number of vertical and horizontal grid
boxes respectively. In each box the potential temperature ($T$) and salinity
($S$) are free variables. The other variables in the problem $\rho$
(density) and $\Psi$ (streamfunction) are dependent on $T$ and $S$.

A variant of the standard procedure (Bryan, 1969) for running an
ocean model is used. The steady state equations with restoring boundary
conditions on temperature and salinity are first solved using Newton's
method and a salinity flux is then deduced. This
flux is then used as the surface condition for salinity and the
equations are solved again with these mixed boundary conditions.
Judicious variation of the initial guess  can be used to find different steady
solutions under those mixed conditions.

It is helpful to arrange the independent variables as a vector ${\bf
x}=(S_{11},S_{12},\dots$ $,S_{21},\dots,S_{mn},T_{11},\dots,T_{mn})$.
Then the advection-diffusion equations for both temperature and salinity
can be written concisely as

$${\bf \dot x}=F({\bf\Psi},K_h,K_v){\bf x},$$
or using summation convention
$$\dot x_i=F_{ij}({\bf\Psi},K_h,K_v)x_j,\ \  i=1,N\eqno(1)$$
where $F$ is a matrix, whose coefficients depend on ${\bf\Psi},K_h$ and
$K_v$, and represents the fluxes due to advection, horizontal and
vertical eddy diffusion, convection and the boundary conditions.
The vertical diffusion coefficient, $K_v$, is to be understood as
including the effects of convection as outlined in the Appendix.
The non-linearities arise through the dependence of $\bf\Psi$ and $K_v$
on $\bf x$ by way of the density. Fortunately $F$ is sparse and
therefore relatively easy to deal with.

For small disturbances, $\delta {\bf x}$, around any state $\bf x$,
(1) can be expanded in a Taylor series,

$$\dot x_i+\delta\dot x_i=F_{ij}x_j+F_{ij}\delta x_j+\left\{ {\p F_{ij}\over
\p \Psi_k}\delta\Psi_k +{\p F_{ij}\over
\p K_v}\delta K_v \right\} x_j+O(|\delta{ \bf x}|^2),\ \  i=1,N\eqno(2)$$
and since $\bf\Psi$ and $K_v$ are functions of $\bf x$, (2) can be written as

$$\dot x_i+\delta\dot x_i=F_{ij}x_j+F_{ij}\delta x_j+\left\{ {\p F_{ij}\over
\p \Psi_k}{\p \Psi_k\over \p x_l} +{\p F_{ij}\over
\p K_v}{\p K_v\over \p x_l} \right\} \delta x_l x_j+O(|\delta {\bf
x}|^2).\ \  i=1,N\eqno(3)$$
The validity of this approach depends crucially upon the fact that
model equations (and specifically the
convection scheme) are differentiable. Collecting together the
perturbation quantities, (3) can be simplified to
$$\dot x_i+\delta\dot x_i=F_{ij}x_j+A_{ij}\delta x_j+O(|\delta {\bf
x}|^2),\ \  i=1,N.\eqno(4)$$
where $A_{ij}=F_{ij}+(\p F_{il}/\p \Psi_k)(\p\Psi_k/ \p x_j) x_l+
(\p F_{il}/ \p K_v)(\p K_v/\p x_j) x_l$.
With the mixed boundary conditions, there is a slight complication
since there is the extra condition that the total salinity content of the
ocean remains constant. This implies that the matrix $A$ as defined
above is singular, (i.e., one equation in (4) is a linear combination of the
others). This just reduces the number of independent variables by one,
so the equations are rewritten by eliminating one of the perturbation
salinity variables. All the following equations then hold.

Equation (4) can be used in two ways. If $\bf x$ is near a steady
state and a closer approximation is needed, then setting the LHS of (4)
to zero and solving for  $\delta \bf x$ gives
$$\delta x_i=-(A^{-1})_{ij}F_{jk}x_k, \eqno(5)$$
which generally gives a better estimate to the steady state. This is
simply Newton's method for
finding roots. Alternatively, if $\bf x$ is a
steady state then
$$\dot x_i=F_{ij} x_j=0,\eqno(6)$$
and (4) can be solved to determine the stability and fastest growing
perturbations (given by the eigenvalues ($\lambda_i$) and eigenvectors
($\mbox{\bf x}_i$) of $A$) of the steady state. It should be noted
that the steady state found under restoring conditions remains a
steady state upon a switch to the derived flux condition (i.e. the
matrix $F$ does not change). However the matrix $A$ does change and
hence so do the stability characteristics. All linear stability
analyses are done under mixed boundary conditions.

Newton's method is the most effective method to use to find a zero of
a non-linear multi-dimensional system of equations if the
differentials can be found analytically (Press {\em et al}, 1990). It
converges
quadratically to the root provided the initial guess is good
enough, but it does have rather poor global convergence and so some
insight into what kind of solution is expected helps, (this can be
gained from a time-stepping model). Most importantly, it will converge to a
steady state {\it regardless} of whether that state is stable or
unstable, unlike a time-stepped model. On occasion, Newton's method
fails to converge, settling instead on a repeating cycle around the
steady state. Usually this can be alleviated by using an initial guess
closer to the desired solution.

Whenever a steady state has to be found numerically it is important to get
as close as possible to the real steady state before performing a
linear stability analysis. The eigenvalues found will correspond more
accurately to those at the steady state the closer one is to that
point. A linear stability analysis away from a steady
point gives information about how neighbouring trajectories behave
(i.e., how the phase space volume is locally evolving) and can
give misleading information about the stability of the (nearby) steady
state. For instance, near a two dimensional saddle point there may be
points where (locally) the phase space volume is reducing in time
and hence a linear stability analysis there would misleadingly
indicate stability.

With knowledge of $A$ and hence the linear stability criteria, we
effectively have a linear model at a tangent to the full model,
valid near a steady state. Hence, if forcing is added to (1) and if
it has a small enough amplitude, the linear equations (4) will
describe the ensuing motion as long as the perturbation remains
small.

Specifically, if the forcing used is similar to white-noise (all
frequencies present), the linearisation allows us to examine the
contribution of each frequency separately. Furthermore, since the
equations decouple if written in terms of the normal modes (the
eigenvectors of $A$), the resonance
response of each normal mode for each forcing frequency can be found.
Take the case of a normal mode $\mbox{\bf x}_i$ with a corresponding complex
eigenvalue $\lambda_i+i\omega_i$.
Given forcing at a particular frequency $\omega$, the equation
governing the evolution of the complex amplitude ($y_i$) of that mode is
$$\dot y_i - (\lambda_i+i\omega_i) y_i = F_i e^{i \omega t}. \eqno(7)$$
The (possibly complex) $F_i$ is the component of the forcing projected
onto this normal mode. If the normal mode is damped then the long term
behaviour of $y_i$ will be governed by the (periodic) particular
solution of this  equation. This solution has a maximum
amplitude (given unit forcing at all frequencies) when
$\omega=\sqrt (\omega_i^2-\lambda_i^2)$, the resonant frequency. If
$|\lambda_i|> |\omega_i|$ then the mode is critically damped and the largest
amplitude response occurs at $\omega=0$. However, if the normal mode
is {\it sub-critical} with $|\lambda_i|< |\omega_i|$ the largest amplitude
response occurs at the resonant frequency. Hence, if white-noise
forcing is used the periodic solution at the resonant frequency would be
the largest significant response.
These periodic solutions are stable if and only if the steady state
around which they oscillate is also stable (i.e. $\lambda_i<0$ for all $i$).
This gives an estimate of the dominant periods of oscillation around
a steady state. It could be that there are no sub-critical modes in which
case the linear response will resemble red noise (high frequencies
damped) with no significant peaks.

Once the linear stability has been established efforts are made to
determine where in parameter space there are changes to the stability.
These are points at which a stable steady state becomes unstable, such as
when the horizontal diffusivity decreases. At these points,
there are bifurcations to different asymptotic forms of behaviour.
There can be branching of solutions leading to different (stable or
unstable) steady states or from a stable state to a limit cycle (a
Hopf bifurcation). This will be discussed further in \S 5.

\head{3. Experiment 1: Variation of the eddy diffusivities}

MSH used the WS  model with low-amplitude stochastic
forcing at the surface to find the flows corresponding to a number of
different values for the eddy
diffusivities. They found century-scale periodic oscillations
around various steady states for typical values of the diffusivities
and large vacillations between steady
states when the horizontal diffusivity was relatively large.
This first experiment repeats the runs performed in MSH but with the
revised convection scheme and the new methodology described in the
previous section.
\newpage
\subhead{3.1 Specification}

The range of values taken for the vertical and
horizontal eddy diffusivities ($K_{v-eddy}$, $K_h$) is the same as in MSH.
Henceforth the vertical eddy diffusivity is written as $K_v$ to
correspond to previous work, but it should not be confused with the
non-constant $K_v$ mentioned in the previous section and Appendix. The
steady state equations for the model are first solved using restoring
boundary conditions on temperature
and salinity with a time constant of 70 days. The salinity
and temperature profiles used in the restoring conditions are the same
as used previously and are an approximation to the
present day climatological averages for the Atlantic. This forces a
two-cell symmetric circulation pattern. From this solution a salinity flux is
derived which is then applied as the surface boundary condition for
the salinity.

Next, the steady state equations are solved with the new boundary
condition and, by choosing a suitable initial guess, a one-cell northern
sinking circulation pattern is found. For completeness, the models
were also led to converge to a southern sinking pattern. However, since
the models were symmetric this was the exact reverse of the
northern sinking pattern. As would be expected the method converged to
the two-cell pattern (even if unstable) if the initial guess was close
or the one-cell pattern did not exist.
The linear stability of the steady states
was determined using the procedure outlined in the previous section.

The resolution, pressure gradient closure parameter $\epsilon$ and time step
were all fixed at 15 horizontal boxes and 9 vertical levels,
0.5 and 14 days respectively as in MSH. The basin stretched from $80^o$N to
$80^o$S, the horizontal eddy diffusivity $K_h$ was one of 1,2,5,10 or 15
($\times10^3$) $\hbox{m}^2 \hbox{s}^{-1}$ while the vertical eddy
diffusivity $K_v$ was chosen from among 1,2,4 and 10 ($\times
0.5\times10^{-4}$) $\hbox{m}^2 \hbox{s}^{-1}$.

\subhead{3.2 Results}

To discuss the results, it will be convenient to refer to each of the
cases by an ordered pair
referring to the vertical and horizontal diffusivities. Thus the
ordered pair (4,5) refers to the case where $K_v=4\times0.5\times10^{-4}$
$\hbox{m}^2 \hbox{s}^{-1}$ and $K_h=5\times10^{3}$ $\hbox{m}^2 \hbox{s}^{-1}$.
The canonical diffusivity case mentioned in MSH corresponds to the
pair (2,1).
Under restoring conditions, the results (outlined in Table 1) show
that in all cases the method converged to a symmetric two-cell circulation
pattern (see Fig. 1) from which the salinity flux at the surface could be
determined.

The results of the linear stability calculation in which the
perturbation field is $\propto \exp(\lambda t)$ (Table 2) show that
except for large $K_h$ the two-cell pattern under mixed boundary
conditions is unstable. Complex eigenvalues imply the existence of slowly
growing or damped oscillatory components. As the horizontal
diffusivity increases, the magnitude of the largest positive
eigenvalue decreases and passes through the origin for $K_h$ between 5
and 10$\times10^{3}$ $\hbox{m}^2 \hbox{s}^{-1}$ and then stays
negative. For these larger values of $K_h$ the two-cell
circulation is {\it stable}. This is confirmed by time-stepping the model
with the salinity flux condition in the high $K_h$ cases and seeing that
the model stays with a two-cell pattern. This implies that
there is a curve in the diffusivity parameter space across which there
is a change of stability and hence a bifurcation to other steady
states. If a close approximation to the critical parameter values can
be found then the bifurcation point could be examined and the nature of
the branching steady states determined.

The eigenvector for a given diffusivity pair corresponding to the
eigenvalue with largest positive real part is the fastest growing
perturbation. Any initial
perturbation to the steady state can be expressed as a linear
combination of all the eigenvectors, but with time the coefficients of
the other eigenvectors will grow more slowly or decay and only the
most unstable eigenvector
will remain (provided the amplitude is sufficiently small so that the
regime is still linear). These unstable eigenvectors for the two-cell
circulation for the different diffusivity pairs are all very similar and are
anti-symmetric in the temperature and salinity
fields. This corresponds to a strengthening of one of the cells over
the other (depending on the sign of the mode coefficient). An example
of the salinity perturbation for the unstable
mode in the case (2,1) is given in Fig. 2 . The temperature field
has much the same structure although it is the salinity field that
dominates the density perturbation. The streamfunction anomaly is
single-signed and thus corresponds to an increase in the circulation
of one cell over the other, (which one depends on the initial
perturbation). Since the two-cell
circulation gives a symmetric distribution for salinity and
temperature (see Fig. 1) this instability is symmetry breaking.

The oscillatory eigenvectors represent modes which all have
symmetric temperature/salinity/density perturbations, as seen for example in
Fig. 3. Periods range between 400 and 700 years and $e$-folding time scales
between 400 to 1000 years. The eigenvectors are complex (${\bf
v}_r\pm i{\bf v}_i$) and the perturbations over a period
follow the pattern ${\bf v}_r \rightarrow -{\bf v}_i \rightarrow -{\bf v}_r
\rightarrow {\bf v}_i \rightarrow {\bf v}_r$ etc. (Note that the
picture on the RHS of Fig. 3 is the imaginary part of the
eigenfunction and therefore corresponds to the perturbation after a
quarter period multiplied by -1). The associated streamfunction
perturbations are antisymmetric and thus correspond to simultaneous
increases (or decreases) in the overturning of the two cells.

In all the cases where the two cell
circulation is unstable there are normal modes that are sub-critical.
If the system is forced, these modes would resonate at century-scale
periods, but since there are unstable modes (some eigenvalues with
positive real part),
these resonant forced periodic solutions are unstable to small
disturbances and would not be realised in practice. There were no
sub-critical modes found in the cases where the two-cell circulation
was stable. As the damping increases (the real part of the
eigenvalue becomes more negative), the structure of the
eigenvectors increases in complexity.

With the new mixed boundary conditions the steady state equations were
solved using a suitable initial condition to coax it to a northern
sinking pattern.
In most of the cases the model converged successfully to such a
one-cell pattern, all of which were similar to the equilibrium for the
case (2,1) shown in
Fig. 4. There is a tendency for a minor southern sinking cell (e.g. in
case (10,10) with a maximum overturning streamfunction of about 1/5 that
of the larger cell) to
appear at the surface as the diffusivities become larger.
The maximum value of the streamfunction found
in each case is given in Table 3, which can be compared with
Table 1 in MSH. In two cases, (4,15) and (10,15), the solution always
converged to the two-cell pattern and we conjecture that there is no
one-cell steady state in these cases, (had the state merely been
unstable, Newton's method should have converged to it).

The flow associated with the slowest decaying mode (the smallest
negative eigenvalue) for the one-cell pattern represents a slowing down or
speeding up of the one-cell circulation without, however, changing the
structure. As noted previously, higher modes have more complex structures.

Cases that are of particular interest here are those that have
sub-critical normal modes. In many cases corresponding to smaller
values of the horizontal diffusivity, there exist just one pair of
sub-critical modes (Table 5). The periods at which they resonate
(included in the table) correspond very well with the dominant
frequencies seen in MSH. In MSH, the dominant period for the
(2,1) canonical diffusivity case was 227 yrs which is close to the 286
yrs predicted here. Similarly, the (1,1) case has a period of 250 yrs
in MSH, which again is just under the 285 yrs here. This implies that
the oscillations around steady states seen in MSH are very close to
being linear.

An example from case (4,1) of the sub-critical oscillatory
perturbation is shown in Figs. 5 and 6 . What is the mechanism for
this oscillation? The flow of the basic state is very similar to that
shown in Fig. 4 so the advection of anomalies by the main flow is
northward at the surface and southward at depth. If we examine the surface
negative salinity anomaly in Fig. 5a, it is clear that a quarter of
a period later it has been subducted into the down-welling zone (Fig.
5b) and that a further quarter of a period later it is entirely
submerged (Fig. 5c). It takes approximately 3 (900 yrs)
periods to go completely around, which is a reasonable overturning time
for a one-basin ocean. As the anomaly is advected
its magnitude decreases after it is down-welled and increases only after
it reaches the surface again. In the absence of forcing, the initial
perturbation has an $e$-folding time of only 50 years (about a quarter
of a period) so this oscillation is not self sustaining. With forcing
at the surface of the right frequency the weak salinity
anomalies at the surface in the South can be amplified as they are
advected northward.

Figure 5 should be compared with Fig. 10 in MSH where salinity
anomalies in the (2,1) case are similarly advected north, down-welled
and diffused in the deep ocean.

The linear perturbations seen here are made up of two separate
effects; the advection by the basic flow of the perturbation
temperature and salinity fields and secondly, the advection by the
perturbation streamfunction field of the basic temperature and
salinity fields. The anomaly in the streamfunction corresponding to
the salinity
anomaly in Fig. 5a is shown in Fig. 6. The perturbation is mostly
confined to the deep ocean where the basic state of the temperature
and salinity is roughly constant. Hence the advection of the basic
field by the anomalous streamfunction plays only a minor r\^ole in the
oscillation.

What does all this tell us about the likely behaviour of the model
under low-amplitude random forcing? If this forcing is assumed to be
temporally and spatially uncorrelated, all modes will be excited more
or less equally. The existence of the purely real
or critically damped complex modes implies that the largest response
of these modes will occur when the forcing frequency tends to zero. Hence
the spectrum of the output of the model will be skewed towards
red-noise. However, in the case where sub-critical oscillatory
modes exist, the largest response is at the non-zero resonant
frequency and so a peak at this frequency above the red noise would be
expected. The structure of the oscillation will then be very close to
that of the sub-critical normal mode. Variation from this pattern
would be expected if the random forcing increased in amplitude so that
the linearisation was no longer valid. Even if the forcing was
low-amplitude, the resonance response could
have a large amplitude and non-linearities would again become important. In
this case,
there is some scope for a weakly non-linear analysis which
would allow amplitude dependence based on a harmonic balance type
approach (see for instance Chapters 5 and 7 of Jordan and Smith
(1977)). This topic will be treated in a future paper.

Interestingly enough, the analysis also shows that for large $K_h$ the
two-cell pattern becomes more stable and that for some cases, e.g.
(1,15) and (2,15),  there are three stable steady states.
Comparing this situation to the large vacillations seen for the (1,15) case in
MSH indicates that in the latter paper the two-cell pattern seems to be an
attractor. It is tempting to suggest that in cases where the two-cell
pattern is unstable, low-amplitude random forcing is unlikely to provoke a
switch between the two one-cell patterns, but where it is {\it stable},
switches between them via the two-cell pattern are more likely, (cf
Fig. 5 and Fig. 6 in MSH). Determining the periods of these highly non-linear
vacillations is obviously beyond the scope of a linear analysis.

\subhead{3.3 Summary}

It is helpful to quickly summarise and compare the results predicted here
with those seen in MSH. However, it must be borne in mind that the
samples of results given in MSH are not complete and that there are
subtle but real differences in the coding of the two models.

The experimental procedure in MSH precluded any stable two-cell
circulations being seen immediately after the switch to the mixed
boundary conditions, (large salinity anomalies were added to the
northernmost cell to encourage convection there). There is, however,
some indirect evidence from the forced solutions that the two-cell
circulation was stable in at least the cases (1,15) and (10,15). This
is also seen in this study.

Only in the case of (10,15) was there no evidence in MSH of a more than
marginally stable one-cell circulation. Here, for this case, no
one-cell circulation could be found.

Where significant century-scale periods were seen in MSH in the
spectra for cases (1,1) and (2,1), namely 250 and 227 yrs, we find
sub-critical modes with periods of
285 yrs and 286 yrs respectively. Overall both studies see a decrease
in century scale periodic behaviour as both $K_h$ and $K_v$ increase.
Generally, the amplitudes of the response of the model used in MSH
increases linearly with the amplitude of the random forcing. This indicates
that the linearised approach taken here is valid.

In the only long time integration performed by MSH (in the (1,15)
case), three stable steady states were clearly seen, precisely as predicted
here.

Differences between the two studies are also worth noting. In some
cases (e.g. (4,1), (4,2) and (10,1)) we predict that century scale
oscillations should occur around the one-cell circulation, although
there does not seem to be such a peak in the spectra for these cases in
MSH. Also, there is clearly a stable one-cell circulation in the case
(4,15) in MSH but we were unable to find it here. Further, we predict
that the two-cell steady state is stable in both the 10 and 15$\times
10^3 \mbox{m}^2 \mbox{s}^{-1} K_h$ cases but as stated above, evidence
for this is only shown for the two cases (1,15) and (10,15).

Possible reasons for these differences are discussed in \S 5.
\head{4. Experiment 2: Variation of the restoring time constants}

The standard boundary condition for temperature used in ocean models
states that the heat flux at the surface, $Q$, is proportional to the
difference between the ocean surface temperature, $T$, and the apparent
atmospheric temperature, $T_a$. The constant of
proportionality can be written as $\Delta z/\tau_T$, where $\Delta z$
is the depth of the first grid point and $\tau_T$ is a restoring time
constant for the temperature. Depending on the derivation of this
boundary condition (assuming an infinite or zero heat capacity
atmosphere) the $\tau_T$ can have values ranging from 50 to 600
days. The apparent atmospheric temperature
takes into account the effects of evaporation and solar radiation but is
close to the actual equilibrium atmospheric temperature almost
everywhere except at the equator (Haney, 1971). Since we are only using an
analytic approximation to the atmospheric temperature, the
distinction need not concern us.
The restoring boundary condition for salinity is generally
written in the same manner.

The experiment carried out below is a simple test of the sensitivity
of the zonally
averaged thermohaline circulation model to changes in the values of
the restoring time constants $\tau_T$ and $\tau_S$. The values chosen
ranged from 50 days to 600 days and the same procedure as in
experiment 1 was followed. Strictly speaking, the apparent atmospheric
variables have different interpretations depending on the derivation
used, but for this simple experiment it will suffice to use the same
form for each.

\subhead{4.1 Specification}

In this experiment the effect of different restoring times for the
salinity and temperature $\tau_S,\tau_T$ on the one-basin circulation
is studied. The values of 50, 300 and 600 days are used for each (see
Table 6) and the same apparent atmospheric temperature or salinity
profiles are used throughout.

The eddy diffusivities and closure parameter used are taken from
Wright and Stocker (1992);  $K_h=1\times 10^3 \hbox{m}^2
\hbox{s}^{-1}$, $K_v=0.5\times 10^{-4} \hbox{m}^2 \hbox{s}^{-1}$ and
$\epsilon=0.45$, while the other factors and
procedures remain as in experiment 1.

\subhead{4.2 Results}

Under restoring boundary conditions, the solution in all cases converged
on the two-cell pattern, as occurred in \S 3, and the salinity flux
was then diagnosed. Generally, the shorter the restoring time, the more
constrained the field. Therefore, since the temperature and salinity
fields are diffusive, a longer restoring
time implies a reduction in the surface gradients of the field.
The THC is driven by the equator-to-pole temperature gradients
but opposed by the salinity gradients, so we would expect a decrease
in the value of the overturning streamfunction as $\tau_T$
increased and an increase as $\tau_S$ increases. This is precisely
what is seen in Table 6 although the variations are relatively
slight. The maximum absolute value of the streamfunction for the two-cell
circulation is not very sensitive to the two parameters.
The structure of the steady state in all cases
is very similar to the example shown in Fig. 1.

A linear stability analysis (Table 7) of the two-cell circulation
does reveal some significant differences between the cases,
particularly when the surface temperature is relatively highly constrained.
There is clearly an increase in stability as the value of $\tau_T$
increases (in good agreement with the results of Zhang {\it
et al}, 1993). However, in contrast with the conclusions of Tziperman {\it
et al} (1994), there is no general increase in
stability with $\tau_S$. For $\tau_T$ of 50 days (Table 7) and for the
cases $\tau_T$ of 30 days and $\tau_S$ of 30 or
120 days (not shown), the results show a {\it decrease} in the
stability with $\tau_S$. This is reversed for larger values of $\tau_T$ and
if the constants are increased together there is a
consistent increase in stability. However, the two-cell circulation never
(for these parameter values) becomes absolutely stable upon a switch
to mixed boundary conditions.

The remarkable increase in the growth rates for the $\tau_T=50$ days
case as $\tau_S$ increases deserves further comment. A further analysis
of this case done with many more values of $\tau_S$ indicated that
there is a smooth increase with $\tau_S$ of the growth rate. However,
the structure of the most unstable normal mode is strikingly
different at the two extremes. Generally, the most unstable mode has
an anti-symmetric structure in the salinity and temperature fields (cf
Fig. 2) and the perturbation is seen over the whole domain.
In contrast, the most unstable modes in
the $\tau_S=$ 600 day case are very localised in the polar regions.
The streamfunction perturbations associated with these represent intense
and localised high latitude up/down-welling. The reason for this lies
in the minor differences that occur in the density structure as $\tau_S$
increases. Since the salinity gradients reduce with $\tau_S$, boxes in
the polar regions have a higher salinity and this serves to actually
reverse the surface density gradient there. The densest surface boxes now
occur one grid box away from the boundary and most convection and
downward advection occurs there (Fig. 7). The top few polar boxes on
the other hand are
now isolated from the main flow and become neutrally stable. Note that
the convection scheme (as described in
the appendix) has the highest gradient with
density at this point. Hence, the linear stability analysis upon a
switch to mixed boundary conditions shows that any positive density
anomaly causes (proportionately) a very large increase in $K_v$, the
diffusion coefficient. Indeed, further tests on these cases show that
there is a high sensitivity of the growth rate with the convection
parameters. This effect is then mainly an artifact of the convection
scheme used but one that does not occur in more
standard cases.

The salinity fluxes derived for the different restoring
times vary only slightly. They decrease in magnitude by a few
percent (mainly at the equator) with both increasing $\tau_S$ and
$\tau_T$. As a consequence the maximum value for
the streamfunction in the final one-cell circulation under mixed
conditions is almost completely insensitive to the restoring time used
for the salinity boundary condition (Table 8). The slight decrease in the
circulation as the temperature restoring time constant increases can
be explained by a weakening of the N-S density gradient caused by a
weaker temperature gradient which occurs because the surface water
does not react as quickly to the atmosphere. There is also little
variation in the largest
eigenvalues for the one-cell steady state as $\tau_S$ increases (Table
9). However, there is a small decrease in the
stability as $\tau_T$ increases. This is to be expected since the
dominant perturbations from the steady state will tend to persist
longer as the restoring time increases.

As would be expected in light
of Table 5, there are sub-critical modes for these cases that would
be expected to resonate given suitable forcing. The results in Table
10 show that, surprisingly, there does seem to be a large dependence
on the value of $\tau_S$ used in the spin up. This indicates that these
higher modes are more sensitive to the structure and magnitude of the
derived salinity flux. As before, the increase in period with $\tau_T$
is due to a weakening of the restoring force on any perturbation.

In conclusion, it seems that this zonally-averaged model has only a
mild sensitivity on the restoring times used. The stabilising
tendencies noted by previous investigators of increased $\tau_T, \tau_S$ seem
to be borne out (except for the one anomalous case due to the highly
delicate nature of the convection scheme). In contrast to the GCM results, in
no case does the solution arrived at under restoring boundary
conditions actually become stable.

\head{5. Conclusions}

A new methodology has been used in this paper to examine the behaviour
of a
one-basin zonally-averaged thermohaline circulation model. The model
is based on the equations and parameterisations introduced by Wright
and Stocker (1991), but modified to include a smooth
(differentiable) convection scheme. Instead of time-stepping the model
to equilibrium as in earlier studies, the steady state equations are
first solved directly using
Newton's method under both restoring and mixed boundary conditions.
This has the advantage of greatly increased
efficiency and achieves much better accuracy in solving the
spatially discretised equations. A much more detailed picture of how
the model behaves in various parameter regions can be built up in a
fraction of the time needed when using the spin-up approach. A linear
perturbation analysis of the steady states then gives
information about the absolute stability of the states, the nature and
structure of the fastest growing modes and what kind of oscillatory
solutions might be expected. Specifically, the period and structure of
any resonance response is readily found. The existence of the multiple
states does not depend upon the preliminary run under restoring conditions
before a salinity flux has been deduced. In fact, if a suitable
salinity flux is prescribed independently, an unstable two-cell
steady state can still be found.

In \S 3, it was shown that this methodology has the ability to
efficiently reproduce and explain results that were previously (and
laboriously) calculated by time stepping the Wright and Stocker model
under low-amplitude stochastic surface forcing (Mysak {\em et al},
1993). There is a
high degree of correspondence between the results of MSH and those
presented here with regard to determining the
absolute stability of each state, the characteristics of each state
and  the resonant century scale oscillations. The minor differences
are most probably explained by a slight decrease in the (numerical)
diffusivity of the present model because of the different convection scheme.
The conclusion drawn is that (in this model at least) the thermohaline
circulation can resonate with century-scale periods around various
steady states and that these oscillations are characterised by the
amplification of surface salinity anomalies and their advection around
the basin.

A comparison of the conclusions made here concerning the bifurcations
of this model can be made with other two-dimensional (non-rotating)
THC models studied by other investigators, specifically Thual and
McWilliams (1992) and Quon and Ghil (1992).
For large values of the diffusion coefficients, the model presented
here has only one steady state, a stable two-cell circulation pattern.
A cross-section through the bifurcation surface is shown in Fig. 8. The
ratio between the horizontal and vertical diffusivities is kept fixed
at $3\times 10^7$ and the vertical diffusivity is increased slowly
from  $0.5\times 10^{-3} \mbox{m}^2 \mbox{s}^{-1}$ to $7.5\times
10^{-3} \mbox{m}^2 \mbox{s}^{-1}$. For small values of the
coefficients there exist two stable one-cell patterns and an unstable
two-cell pattern. As the coefficients increase the two-cell
circulation pattern becomes stable, and as they increase further the one-cell
circulation patterns disappear. It is likely that there is a
pitchfork bifurcation as the stability of the two-cell pattern changes
which connects the paths; however
no evidence was found for these intermediate unstable steady states.
In the Thual and McWilliams model there is a very similar bifurcation
pattern seen (Figure 3 in their paper) but with an
additional salinity-dominated steady state not present here. A
symmetric temperature dominant steady state becomes unstable at some
critical parameter and bifurcates to a two mutually opposite one-cell
circulations which only exist for a limited parameter range.
A similar pattern is also seen in the Quon and Ghil model. As the
thermal Raleigh number
(inversely proportional to the temperature diffusion coefficient)
increases, a symmetric two-cell circulation bifurcates {\it smoothly}
to an increasingly asymmetric one-cell dominant pattern (see their
Figures 14 and 16). Hence, it seems that the bifurcation structure seen in
two-dimensional THC models is quite robust.

The work in experiment 2 dealing with different restoring times for the
temperature and salinity relaxation boundary conditions indicates that
the circulation changes very little as the time constants change over
an order of magnitude. One possible explanation for the disparity
between this result and that of Tziperman {\it et al} (1994)
is that
changes in the restoring times will likely have an effect on three
dimensional processes with
shorter time scales (i.e., over a few years) which are not represented
in the Wright and Stocker model. Hence any changes of stability found
in GCMs as $\tau_T$ and $\tau_S$ vary, are unlikely to be found in this
model. This implies the rather welcome conclusion that the eventual
steady states or long term behaviour of the 2-D model is not crucially
dependent on the spin up process or on the exact formulation of the
relaxation boundary conditions. In one case with a large $\tau_S$
but small $\tau_T$, there was an anomalously large growth rate for
perturbations to the two-cell circulation. This was due to a small pool
of near-neutrally stable water collecting at the poles which could be
triggered to start convecting with only a small salinity perturbation.
This subsequent large linear growth rate is hence very sensitive to
the details of the convection scheme. Note that for large growth rates
the time for which the linear approximation is valid will be much
smaller than in other cases.

Other limitations of this work should also be mentioned. Most interesting
behaviour is non-linear (large amplitude), and if a linear analysis is
to provide any kind of qualitative indications of asymptotic behaviour it
must be complete, i.e. {\it all} the steady states should be found and
an attempt made to describe the phase paths.
However, there is no certain way of knowing whether that point has
been reached. Also, if higher amplitude forcing is used,
then the linear modes are only useful as a first approximation to the
non-linear resonance response and give no information about the much
more rich behaviour of even a weakly non-linear model. There is some
scope for analytical results in the weakly non-linear regime and this
will be discussed in a future paper.
Notwithstanding these caveats, the methodology presented here has
proved to be an efficient way of
describing the behaviour of a 2-D
model. It would be interesting in future work to examine the stability
of the solutions when there
are two or more basins connected by an Antarctic Circumpolar Current.
In particular,
are the century scale oscillations still prevalent? The effect of
relative basin widths could also be clearly examined. It
could also be worthwhile to try and apply the same methodology to
three dimensional models.

\head{6. Acknowledgments}

This work was supported by research grants awarded to LAM from NSERC
and Fonds FCAR. The authors are grateful for the helpful comments
received from two anonymous referees.

\head{Appendix A: Derivation of a differentiable convection scheme}

As an alternative to the explicit mixing algorithm used by Wright and
Stocker (1991), an attempt is made to determine the convective flux
produced by their scheme and translate it into an equivalent diffusive
flux which would vary smoothly with the densities of the grid boxes.
The basic idea can be made clear by considering a column of
two overlying boxes with fluid of different densities.

Assume that the fluid at the interface is unstably stratified,
hence a convective correction must be made. According to the explicit
mixing scheme, the potential temperature and salinity of the two boxes will be
averaged and each box will now have identical characteristics.
For each box let $\Delta z_i, T_i$ and $S_i$ be the (fixed) depth of
the box, the box-averaged potential temperature and the box-averaged salinity
respectively, where $i=1,2$. The new $T$ and $S$ for the boxes can
then be written as
$$ T={\Delta z_1T_1+\Delta z_2T_2 \over \Delta z_1+\Delta z_2},\ \
S={\Delta z_1S_1+\Delta z_2S_2 \over \Delta z_1+\Delta z_2}.$$
This implies heat and salt transports between the boxes
{\it over one time step} of
$$M_T={\Delta z_1\Delta z_2(T_2-T_1) \over \Delta z_1+\Delta z_2},$$
and
$$M_S={\Delta z_1\Delta z_2(S_2-S_1) \over \Delta z_1+\Delta z_2}.$$

If there were solely a diffusive flux, using a second order
flux-conservative finite difference scheme (specifically the scheme
outlined by Fiadiero and Veronis (1977) and generalised by Wright
(1992)) would imply vertical fluxes of

$$F_T={2K (T_2-T_1)\over \Delta z_1+\Delta z_2},$$
and
$$F_S={2K (S_2-S_1)\over \Delta z_1+\Delta z_2},$$
through the interface.

The effective diffusion coefficient of the convective scheme is
then determined by equating the two expressions for the fluxes after
taking account of the time step, i.e.

$$K_{conv}={\Delta z_1\Delta z_2 \over 2\Delta t},\eqno(A.1)$$
This is the diffusion that would have been required to operate over
the time step specified to give the same effect as the explicit
convection scheme.

It is also possible to find similar expressions in the case where
there are three boxes in a column. For this situation the effective diffusion
coefficients at the two boundaries between the cells are

$$K_{conv}^1={\Delta z_1\Delta z_2  \over 2\Delta t}{(\Delta
z_1+\Delta z_2) \over (\Delta z_1+\Delta z_2+ \Delta z_3)},\eqno(A.2a)$$
$$K_{conv}^2={\Delta z_2\Delta z_3  \over 2\Delta t}{(\Delta
z_2+\Delta z_3) \over (\Delta z_1+\Delta z_2+ \Delta z_3)},\eqno(A.2b)$$
for boxes 1 and 2 and boxes 2 and 3 respectively.

For more than three boxes, there is no equivalence of convection  and
diffusion since the explicit convection scheme takes into account all the
values of the temperature and salinity over the column. Such non-local
effects cannot be modelled using only a second order scheme.
However the form of the effective diffusion in the two and three box
cases suggests that a diffusion coefficient proportional to the depth
of two neighbouring boxes and inversely proportional to the time step
might produce similar results. For example, at the boundary between
the $i$th and $i+1$th boxes, the diffusion coefficient could take the
form

$$K_{conv}^i=\lambda {\Delta z_i\Delta z_{i+1} \over \Delta t }.
\eqno(A.3)$$
In the two box case $\lambda=1/2$, and in the three box case
$\lambda\approx 1/3$ (depending on the relative sizes of the boxes).
The key point is
to choose a value for lambda that gives the most consistent results.
There are some restrictions on the value of $\lambda$ since numerical
instability will result if the diffusion is too large. There is also a
lower limit on the effective diffusion since it must be
greater than the eddy diffusion coefficient being used, probably by
an order of magnitude. If $\lambda$ is taken to be 0.5 in the full
model then the numerical scheme is unstable. Various experiments with
different values lead to the conclusion that the maximum $\lambda$
consistent with stability  over a range of time steps
and grid spacings is $\lambda=1/3$. The results from using this
scheme and the explicit scheme in the model are qualitatively and
quantatively similar and hence it is used exclusively in the examples in
the text.

The vertical diffusion coefficient used at any interface now depends
in a consistent way on the static stability criteria (the sign of
$\rho_z$ ) at that interface. There is still a discontinuity at the
point where  $\rho_z=0$ which has to be smoothed out if a fully
differentiable scheme is required. Physically this can be justified by
considering the unresolved variations in density on
both sides of the interface. It is entirely probable that the
conditions for convection will occur at points on the interface
without the average cell densities being statically unstable. Hence
the flux of heat and salt will increase over that generated solely by eddy
diffusion ($K_{v-eddy}$) as neutral stability is approached. Therefore, we have
chosen to use a tanh-like functional form for the vertical diffusion
i.e.

$$K_v^i=K_{v-eddy}\exp\left( {1\over 2} \log\left({K_{conv}^i\over K_{v-eddy}
}\right) (\tanh( \gamma( \rho_i -\rho_{i+1}))+1)\right)\eqno(A.4)$$
where the parameter $\gamma$ determines how steep the smoothing is.
This form for the vertical diffusion has the properties that it tends
to $K_{v-eddy}$ for very stable conditions and to $K_{conv}$ as static
instability occurs. Exponential functions are included since the
values of $K_{v-eddy}$ and $K_{conv}$ can differ by orders of magnitude
and with a simple linear smoothing $K_v$ increases much too rapidly.
The results using this form are robust as long as
$\gamma$ is sufficiently large (greater than about $30/\rho_0
\ \mbox{m}^3 \mbox{kg}^{-1}$ ). A uniform value of $50/\rho_0
\ \mbox{m}^3 \mbox{kg}^{-1}$ is used throughout this paper.

There seems to be a contradiction in using this scheme to solve the
steady state equations because of the explicit dependence on the
time-step $\Delta t$ in (A.3). However, this has the effect of
ensuring that the {\em flux} associated with the convection does not
depend on $\Delta t$ and that this flux only depends on the density
differences. There is a plausible argument that says that this
time constant should be associated with the time scales of actual
convective events. Experiments with
various reasonable $\Delta t$ values show very
little difference in the steady states found as $\Delta t$ varied from
1 day to 2 months ($\Delta t$ in time-stepped model runs is usually
about 14 days).

\head{References}
\begin{description}
\frenchspacing
\itemsep=-2pt
\parsep=0pt

\item Bryan, F. 1986.
 High latitude salinity effects and interhemispheric thermohaline
  circulation.
 {\em Nature}, {\bf 323}, 301--304.

\item
Bryan, K. 1969.
 A numerical method for the study of the circulation of the world
  ocean.
 {\em J. Comput. Phys.}, {\bf 4}, 347--376.

\item
Cessi, P. 1994.
 A simple box model of stochastically forced thermohaline flow.
 {\em J. Phys. Oceanogr.}, {\bf 24}, 1911--1920.

\item
Fiadiero, M.~E. and Veronis, G. 1977.
 On weighted-mean schemes for the finite-difference approximation to
  the advection-diffusion equation.
 {\em Tellus}, {\bf 29}, 512--522.

\item
Haney, R.~L. 1971.
 Surface thermal boundary condition for ocean circulation models.
 {\em J. Phys. Oceanogr.}, {\bf 1}, 241--248.

\item
Jordan, D.~W. and Smith, P. 1977.
 {\em Non-linear Ordinary Differential Equations}.
 Oxford Applied Mathematics and Computing Science Series. Oxford
  University Press.

\item
Maas, L. R.~M. 1994.
 A simple model for the three-dimensional, thermally and wind driven
  ocean circulation.
 {\em Tellus}, {\bf 46A}, 671--680.

\item
Manabe, S. and Stouffer, R.~J. 1988.
 Two stable equilibria of a coupled ocean atmosphere model.
 {\em J. Climate}, {\bf 1}, 841--866.

\item
Marotzke, J. 1991.
 Influence of convective adjustment on the stability of the
  thermohaline circulation.
 {\em J. Phys. Oceanogr.}, {\bf 21}, 903--907.

\item
Mysak, L.~A., Stocker, T.~F., and Huang, F. 1993.
 Century-scale variability in a randomly forced, two-dimensional
  thermohaline ocean circulation model.
 {\em Climate Dynamics}, {\bf 8}, 103--116.

\item
Press, W.~H., Flannery, B.~P., Teukolsky, S.~A., and Vetterling, W.~T. 1990.
 {\em Numerical Recipies}.
 Cambridge University Press.

\item
Quon, C. and Ghil, M. 1992.
 Multiple equilibria in thermosolutal convection due to salt-flux
  boundary conditions.
 {\em J. Fluid Mech.}, {\bf 245}, 449--483.

\item
Schopf, P.~S. 1983.
 On equatorial waves and {El Ni\~no. II}: Effects of air-sea thermal
  coupling.
 {\em J. Phys. Oceanogr.}, {\bf 13}, 1878--1893.

\item
Stocker, T.~F. and Wright, D.~G. 1991.
 A zonally averaged ocean model for the thermohaline circulation.
  {Part II}: Interocean circulation in the {P}acific-{A}tlantic basin system.
 {\em J. Phys. Oceanogr.}, {\bf 21}, 1725--1739.

\item
Stocker, T.~F., Wright, D.~G., and Mysak, L.~A. 1992.
 A zonally averaged coupled ocean-atmosphere model for paleoclimatic
  studies.
 {\em J. Climate.}, {\bf 5}, 773--797.

\item
Stommel, H. 1961.
 Thermohaline convection with two stable regimes of flow.
 {\em Tellus}, {\bf 13}, 224--230.

\item
Thual, O. and McWilliams, J.~C. 1992.
 The catastrophe structure of thermohaline convection in a
  two-dimensional fluid model and a comparison with low-order box models.
 {\em Geophys. Astrophys. Fluid Dynamics}, {\bf 64}, 67--95.

\item
Tziperman, E., Toggweiler, J.~R., Feliks, Y., and Bryan, K. 1994.
 Instability of the thermohaline circulation with respect to mixed
  boundary conditions: Is it really a problem for realistic models?
 {\em J. Phys. Oceanogr.}, {\bf 24}, 217--232.

\item
Weaver, A.~J. and Hughes, T. 1992.
 Stability and variability of the thermohaline circulation and its
  link to climate.
 {\em Trends in Phys. Oceanogr., Council of Scientific Research
  Integration, Trivandrum, India}.

\item
Weaver, A.~J. and Sarachik, E.~S. 1991.
 The role of mixed boundary conditions in numerical models of the
  ocean's climate.
 {\em J. Phys. Oceanogr.}, {\bf 21}, 1470--1493.

\item
Wright, D.~G. 1992.
 Finite difference approximations to the advection-diffusion equation.
 {\em Tellus}, {\bf 44A}, 261--269.

\item
Wright, D.~G. and Stocker, T.~F. 1991.
 A zonally averaged ocean model for the thermohaline circulation.
  {Part I}: Model development and flow dynamics.
 {\em J. Phys. Oceanogr.}, {\bf 21}, 1713--1724.

\item
Wright, D.~G. and Stocker, T.~F. 1992.
 Sensitivities of a zonally averaged global ocean circulation model.
 {\em J. Geophys. Res.}, {\bf 97}, 12707--12730.

\item
Zhang, S., Greatbatch, R.~J., and Lin, C.~A. 1993.
 A re-examination of the {P}olar {H}alocline {C}atastrophe and
  implications for coupled ocean-atmosphere modelling.
 {\em J. Phys. Oceanogr.}, {\bf 23}, 287--299.

\end{description}
\newpage

\tbl{|c|ccccc|}{
$K_v$  & \multicolumn{5}{c|}{$K_h$ Horizontal diffusivity} \\
Vertical diffusivity  & \multicolumn{5}{c|}{($\times10^3$
$\hbox{m}^2 \hbox{s}^{-1}$)} \\
($\times0.5\times10^{-4}$ $\hbox{m}^2 \hbox{s}^{-1}$) &
\multicolumn{5}{c|}{} \\
 &1&2&5&10&15\\ \hline
1& 5.95& 5.59& 4.97& 4.41& 3.94 \\
2& 9.03& 9.14& 8.35& 7.17& 6.48\\
4& 14.2& 13.9& 13.3& 11.8 & 10.5 \\
10& 25.3& 25.2& 24.5& 22.7& 20.5 \\}
{1}{The maximum absolute
value for the stream function (in Sv) at steady state for each pair of
diffusivities under restoring boundary conditions. This corresponds to a
two-cell circulation pattern. }

\tbl{|c|ccccc|}{
$K_v$  & \multicolumn{5}{c|}{$K_h$ Horizontal diffusivity} \\
Vertical diffusivity  & \multicolumn{5}{c|}{($\times10^3$
$\hbox{m}^2 \hbox{s}^{-1}$)} \\
($\times0.5\times10^{-4}$ $\hbox{m}^2 \hbox{s}^{-1}$) &
\multicolumn{5}{c|}{} \\
 &1&2&5&10&15\\ \hline
1  &2.20            & 1.87            &1.09     & -0.02    & -0.30 \\
   &$0.75\pm i0.87$ & $0.39\pm i1.24$ &$-0.73\pm i1.71$ & -0.34 &-0.95
\\ \hline
2  &1.77            &1.62     	&0.95     & -0.09    & -0.46\\
   &$0.27\pm i1.47$&$0.07\pm i1.68$ &-0.44 & -0.46 &-1.07
 \\ \hline
4  &1.24            &1.03   &0.50     &-0.38     & -0.38  \\
   &-0.38           &-0.40  &-0.41    &-0.41     &-1.28
\\ \hline
10 &0.66            &0.53   &0.13     & -0.68    & -0.69 \\
   &-0.56           &-0.57  &-0.51    & -0.69    &-1.55  \\}
{2}{The values of the two largest eigenvalues (in $\hbox{(100
year)}^{-1}$) in
the linear stability analysis of the two-cell circulation pattern
under mixed boundary conditions. Positive (negative) values indicate
an unstable (stable) state and complex values indicate
growing or damped oscillations. As both $K_v$ and $K_h$ increase, the two-cell
state becomes more stable.}

\tbl{|c|ccccc|}
{$K_v$  & \multicolumn{5}{c|}{$K_h$ Horizontal diffusivity} \\
Vertical diffusivity  & \multicolumn{5}{c|}{($\times10^3$
$\hbox{m}^2 \hbox{s}^{-1}$)} \\
($\times0.5\times10^{-4}$ $\hbox{m}^2 \hbox{s}^{-1}$) &
\multicolumn{5}{c|}{} \\
 &1&2&5&10&15\\ \hline
1& 11.6& 11.1& 9.8& 8.2& 6.3 \\
2& 18.0& 17.4 & 15.8& 13.5 & 10.5\\
4& 26.7 & 26.1 & 24.3 &21.3 & - \\
10& 45.0 & 44.0& 40.9 & 35.3 & - \\}
{3}{The maximum absolute
value for the stream function (in Sv) at steady state for each pair of
diffusivities under the mixed boundary conditions (cf. Table 1 in
MSH). These results are for a
one-cell northern-sinking circulation pattern. No values are given for the
(4,15) or (10,15) cases because these always converged to the two-cell pattern
as described in the text.}

\tbl{|c|ccccc|}{
$K_v$  & \multicolumn{5}{c|}{$K_h$ Horizontal diffusivity} \\
Vertical diffusivity  & \multicolumn{5}{c|}{($10^3$
$\hbox{m}^2 \hbox{s}^{-1}$)} \\
($0.5\times10^{-4}$ $\hbox{m}^2 \hbox{s}^{-1}$) &
\multicolumn{5}{c|}{} \\
 &1&2&5&10&15\\ \hline
1  &-0.09           & -0.10          & -0.11       &-0.09 &-0.17  \\
   &$-0.38\pm i0.20$&$-0.45\pm i0.17$& -0.44       &-0.36  &-0.70\\
\hline
2  & -0.15           &-0.16          &-0.17          &-0.17 & -0.18\\
   &$-0.53\pm i0.32$&$-0.62\pm i0.32$&$-0.86\pm i0.25$& -0.70 & -0.62 \\
\hline
4  & -0.24           & -0.24          & -0.24        & -0.27     &  -  \\
  & $-0.70\pm i0.49$& $-0.78\pm i0.49$& -0.94   &$-1.32\pm i0.47$& -
\\
\hline
10 & -0.48           & -0.47           & -0.44 & -0.43& - \\
   & $-1.35\pm i0.58$&$-1.22\pm i0.60$ & $-1.48\pm i0.79$&$-1.49\pm i0.68$& -
\\}
{4}{The value of the two largest eigenvalues (in $\hbox{(100
year)}^{-1}$) in
the linear stability analysis of the one-cell circulation pattern.
All values are have negative real parts hence all the states examined
are stable but complex values indicate
(damped) oscillations. No values are given for the(4,15), (10,15) cases since
no estimate of the one-cell circulation could be found.}

\tbl{|c|ccccc|}{\multicolumn{6}{|c|}{Frequencies of sub-critical modes and
their resonant periods} \\ \hline
$K_v$  & \multicolumn{5}{c|}{$K_h$ Horizontal diffusivity} \\
Vertical diffusivity  & \multicolumn{5}{c|}{($10^3$
$\hbox{m}^2 \hbox{s}^{-1}$)} \\
($0.5\times10^{-4}$ $\hbox{m}^2 \hbox{s}^{-1}$) &
\multicolumn{5}{c|}{} \\
 &1&2&5&10&15\\ \hline
1  & $-1.12 \pm i2.47 $ & $-1.21\pm i2.14$ & - & - &$-6.49\pm i7.10$ \\
   & 285 yrs  &  353 yrs&    &     & 218 yrs  \\ \hline
2  & $-1.75 \pm i2.80 $  & $-1.77 \pm i2.49 $ 	& -    & -    & -\\
   & 286 yrs & 359 yrs   &    &     &  \\ \hline
4  & $-2.34 \pm i 3.16 $ & $-2.42\pm i 2.84 $	& -     &-     & -  \\
   & 296 yrs&  423 yrs &    &     &  \\ \hline
10 & $-3.41 \pm i 3.61 $ & -   		& -     & -    & - \\
   & 526  yrs       &    &     &     &  \\}
{5}{For the one-cell circulation, some cases associated with smaller
eddy diffusivities have normal modes that are {\it sub-critical},
i.e., they are not critically
damped and hence have resonant frequencies. These frequencies are
those that would be expected to dominate the response of the system
under white-noise forcing. The century-scale periods for cases (1,1)
and (2,1) compare favorably with those seen in the model runs of MSH.}
\tbl{|c|ccc|}{
$\tau_T$  & \multicolumn{3}{c|}{$\tau_S$ Salinity restoring } \\
Temperature restoring time & \multicolumn{3}{c|}{time (days)} \\
(days) & \multicolumn{3}{c|}{} \\
 & 50 & 300 & 600 \\ \hline
50& 5.71 &5.98 & 6.45 \\
300& 5.68 &5.84 & 6.03\\
600& 5.57 &5.71 & 5.87 \\}
{6}{The maximum absolute
value for the stream function (in Sv) at steady state for each pair of
restoring time values under restoring conditions. This corresponds to a
two-cell circulation pattern. There is a slight tendency for the
circulation to increase with $\tau_S$ and to decrease with $\tau_T$
but both the pattern and magnitude of the circulation are relatively stable. }
\tbl{|c|ccc|}{
$\tau_T$  & \multicolumn{3}{c|}{$\tau_S$ Salinity restoring } \\
Temperature restoring time & \multicolumn{3}{c|}{time (days)} \\
(days) & \multicolumn{3}{c|}{} \\
 & 50 & 300 & 600 \\ \hline
50  &2.17 &  18.9 & 44.4 \\
   &$0.78\pm i0.76$ &16.8 & 44.4 \\ \hline
300  & 0.84 & 0.65  & 0.44 \\
 & -0.11    & -0.11 & -0.11\\ \hline
600  & 0.44 & 0.28 & 0.13 \\
     &-0.07 & -0.07 & -0.07 \\}
{7}{The
value of the two largest eigenvalues (in $\hbox{(100 year)}^{-1}$) in
the linear stability analysis of the two-cell circulation pattern.
Positive values indicate an unstable state and complex values indicate
(growing or damped) oscillations.}
\tbl{|c|ccc|}{
$\tau_T$  & \multicolumn{3}{c|}{$\tau_S$ Salinity restoring } \\
Temperature restoring time & \multicolumn{3}{c|}{time (days)} \\
(days) & \multicolumn{3}{c|}{} \\
 &50&300&600\\ \hline
50 & 11.2 & 11.1 & 10.9 \\
300& 10.8 & 10.8 & 10.7 \\
600& 10.4 & 10.3 & 10.3 \\}
{8}{The maximum absolute
value for the stream function (in Sv) at the steady state under the
mixed boundary conditions. This corresponds to a
one-cell circulation pattern. There is a slight tendency for the
circulation to decrease with $\tau_T$ and $\tau_S$ but as for the
two-cell circulation the pattern and magnitude of the circulation seem
relatively stable. }

\tbl{|c|ccc|}{
$\tau_T$  & \multicolumn{3}{c|}{$\tau_S$ Salinity restoring } \\
Temperature restoring time & \multicolumn{3}{c|}{time (days)} \\
(days) & \multicolumn{3}{c|}{} \\
 & 50 & 300 & 600 \\ \hline
50 &-0.10 & -0.10 & -0.10 \\
300&-0.06 & -0.06 & -0.06 \\
600&-0.05 & -0.05 & -0.05 \\}
{9}{The
value of the largest eigenvalues (in ($\hbox{(100 year)}^{-1}$) in
the linear stability analysis of the one-cell circulation pattern.
There is almost no variation with the $\tau_S$ used in the spin up;
however, as $\tau_T$ increases the one-cell state becomes slightly
less stable due to the reduction in the surface restoring force.}

\tbl{|c|ccc|}{
$\tau_T$  & \multicolumn{3}{c|}{$\tau_S$ Salinity restoring } \\
Temperature restoring time & \multicolumn{3}{c|}{time (days)} \\
(days) & \multicolumn{3}{c|}{} \\
 & 50  & 300 & 600 \\ \hline
50 & 300 & 320 & 390 \\
300& 520 & 700 &  - \\
600& 1900 & -  & - \\}
{10}{The period (in years) associated with the sub-critical normal mode
at resonance. In contrast to table 9, there is a relatively large
variation in the resonant periods with $\tau_S$. This indicates a higher
degree of sensitivity to the salinity flux boundary condition. The
increase of period with $\tau_T$ is to be expected due to the reduction in
the strength of the restoring condition.}
\clearpage
\renewcommand{\baselinestretch}{1.0}\small
\newpage
\begin{figure}[h]
\begin{picture}(400,500)
\put(-72,-50){\includegraphics{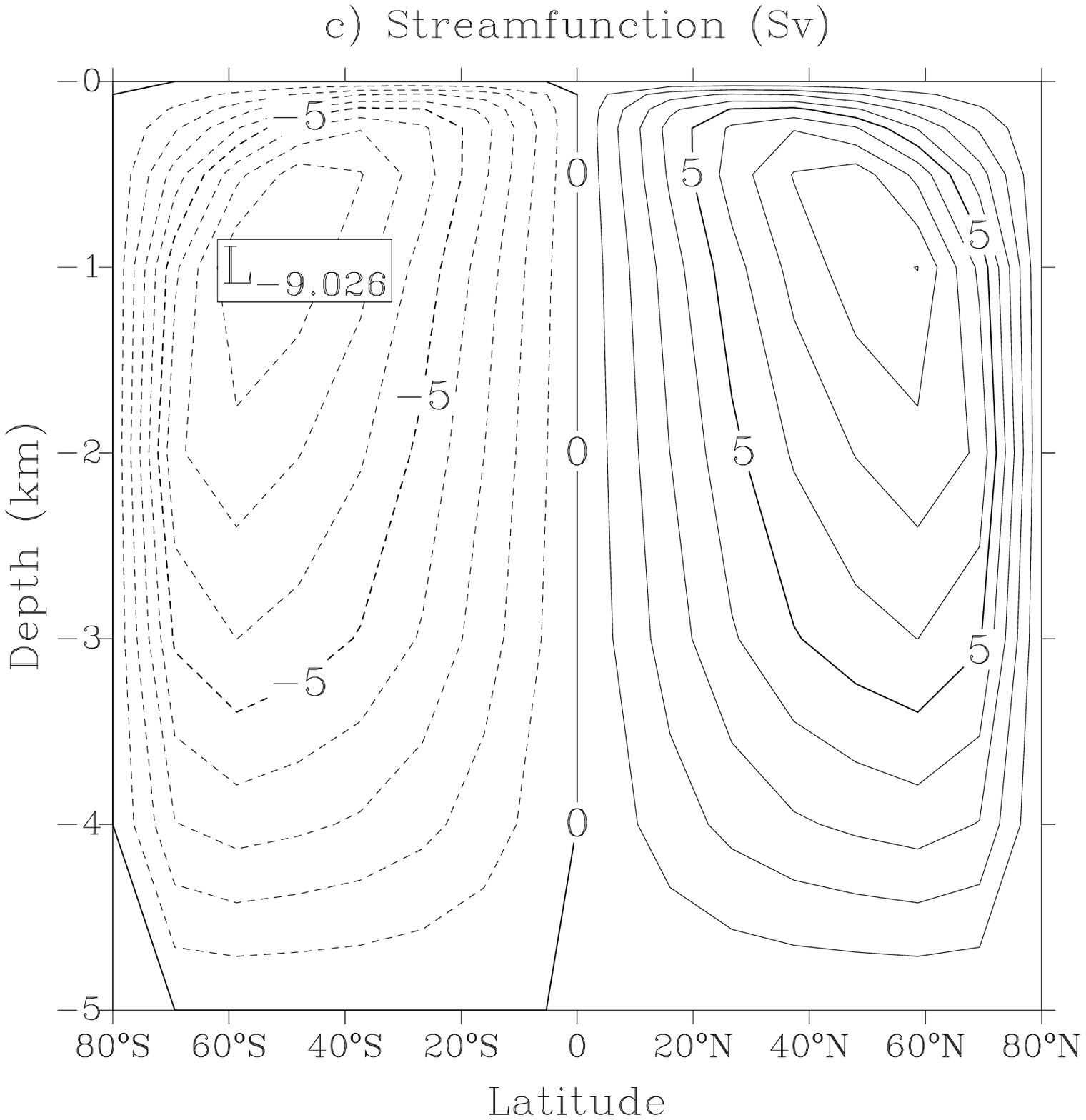}}
\put(-72,200){\includegraphics{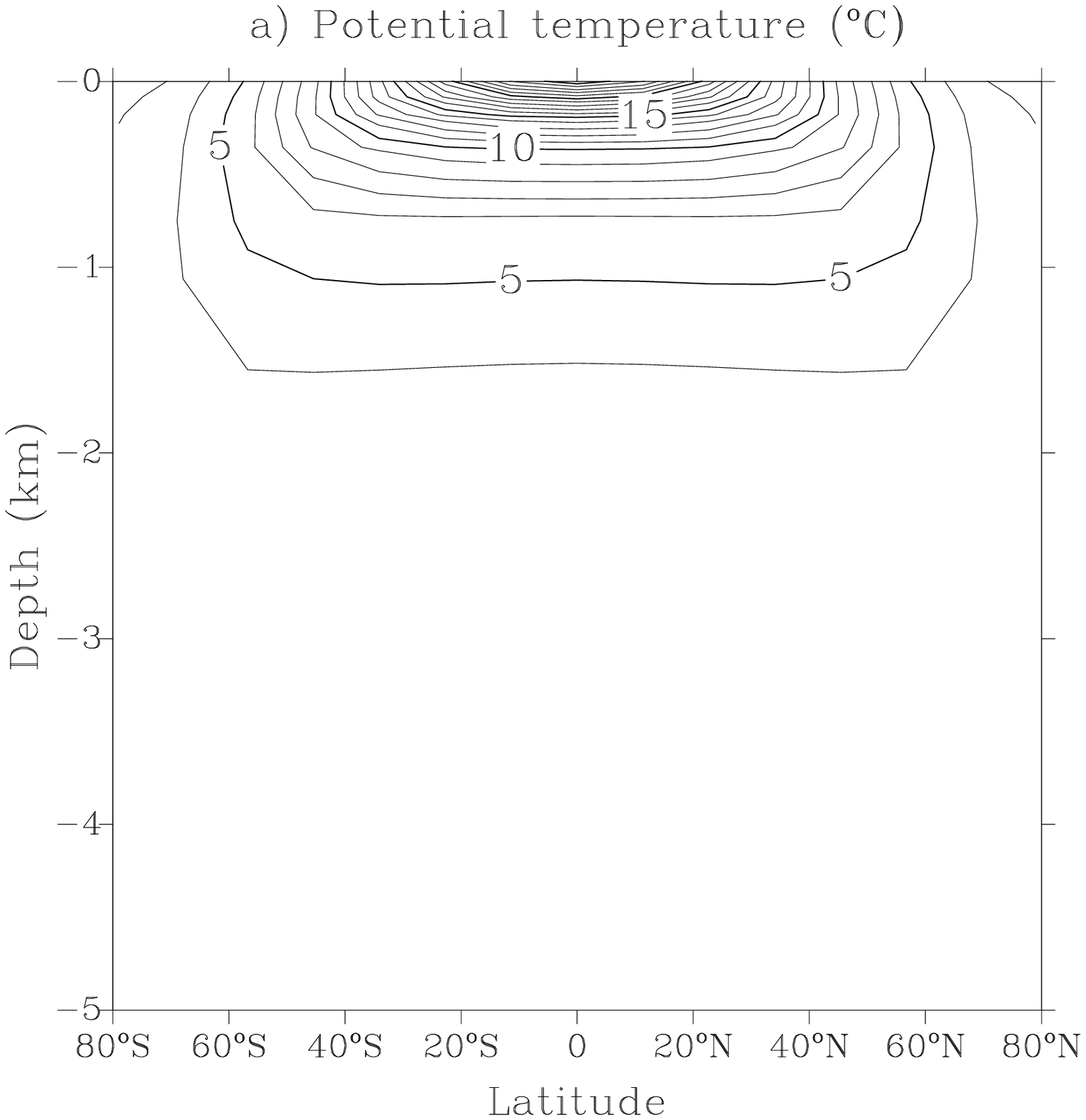}}
\put(200,200){\includegraphics{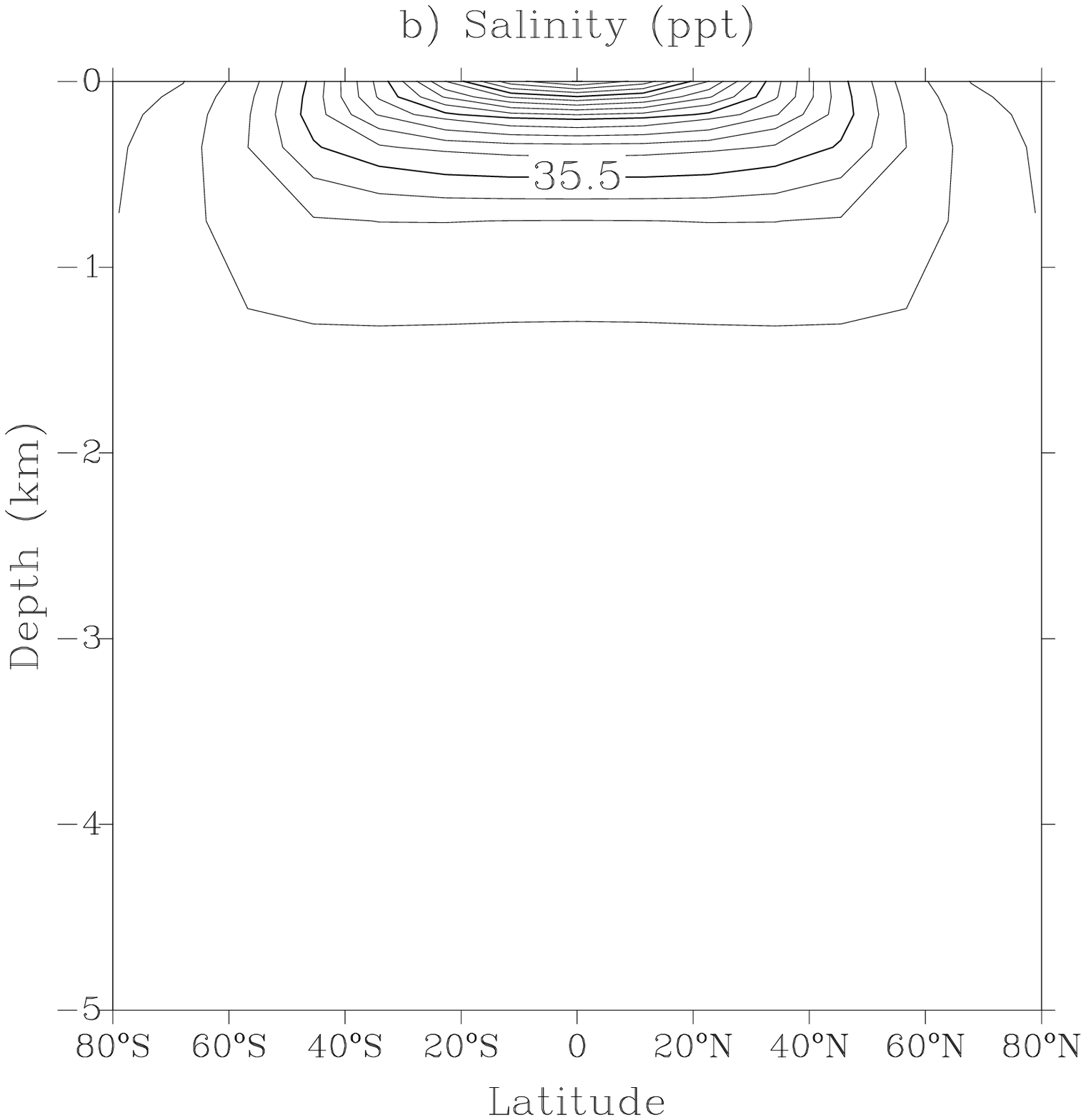}}
\put(194,-50){\includegraphics{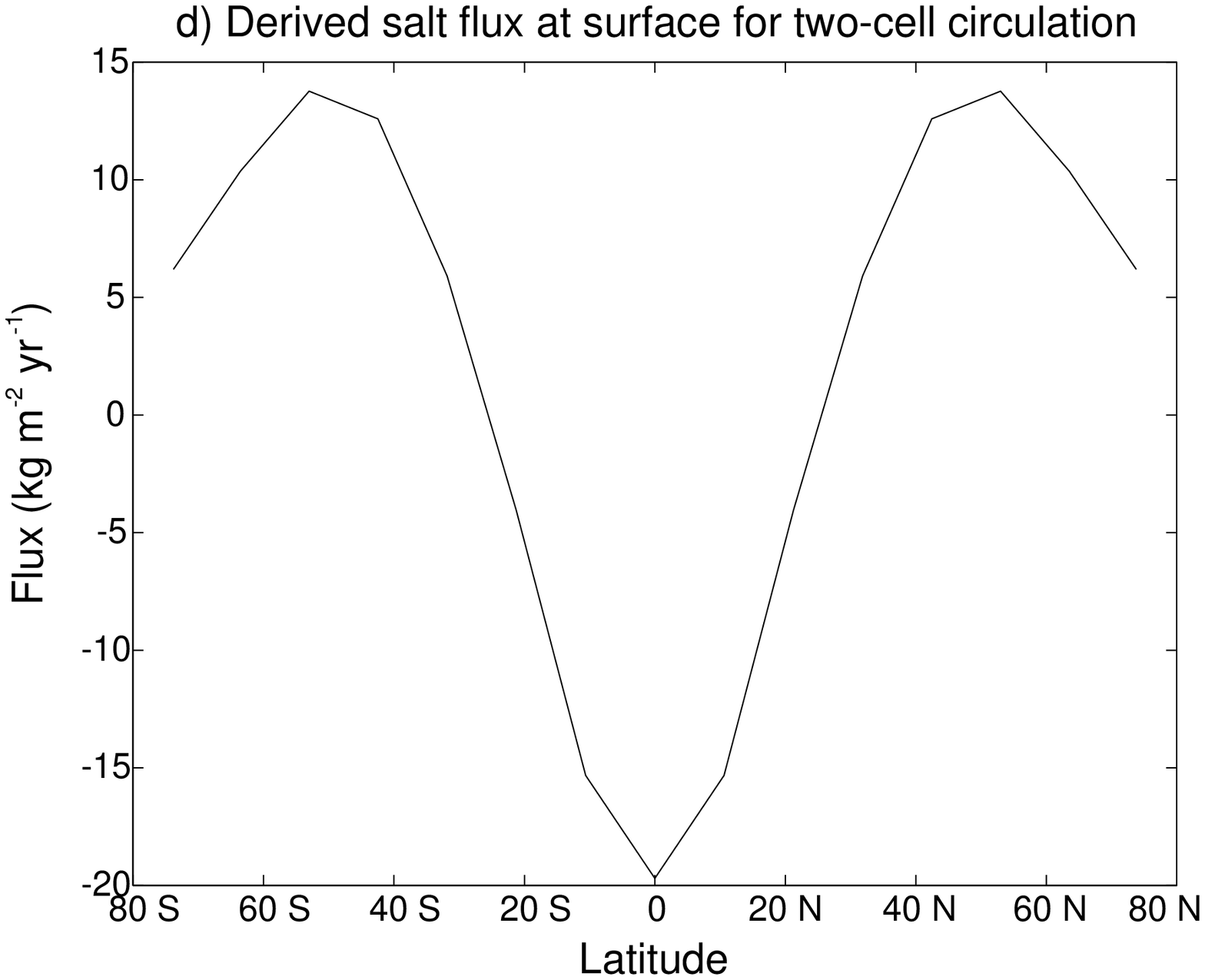}}
\end{picture}
\begin{quote}{\bf
Figure 1}: The (a) Temperature, (b) salinity (contour interval 0.1 ppt), (c)
streamfunction and (d) derived salinity flux for the two cell circulation
in the (2,1) (the so-called canonical diffusivities) case. This
symmetric circulation is reached after solving the equations under restoring
boundary conditions.
\end{quote}\end{figure}\renewcommand{\baselinestretch}{1.5}\normalsize

\fig{400,300}{48,0}{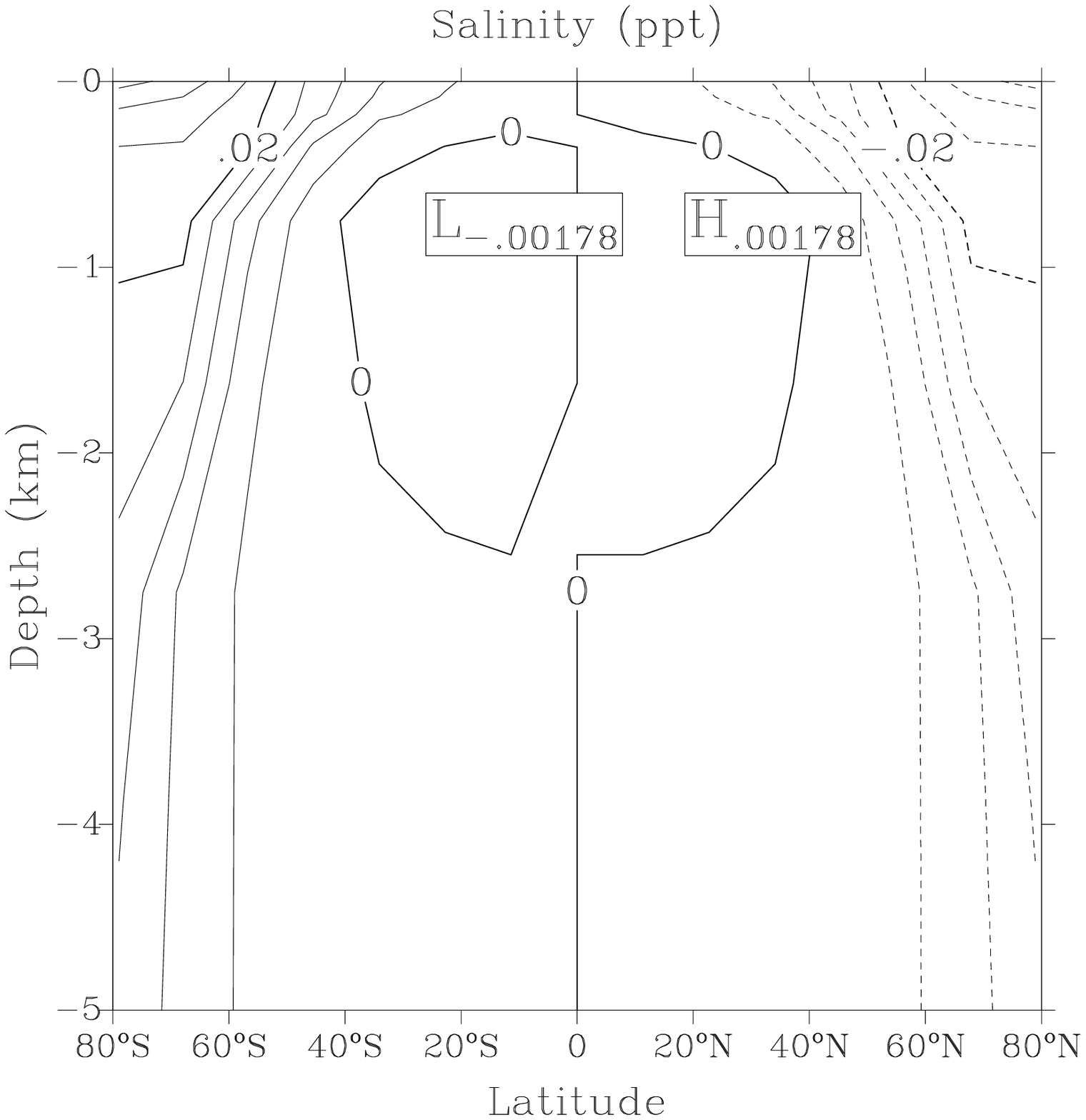}{50.0}{The salinity perturbation that
corresponds to the fastest growing (non-oscillatory) mode for
the two-cell circulation in the (2,1) (canonical diffusivities) case.
This mode is anti-symmetric and thus causes one cell to become larger
than the other which finally leads to the one-cell circulation seen in
Fig. 4. The corresponding temperature perturbation is very similar.}

\renewcommand{\baselinestretch}{1.0}\small
\begin{figure}[h]
\begin{picture}(400,300)
\put(-72,0){\includegraphics{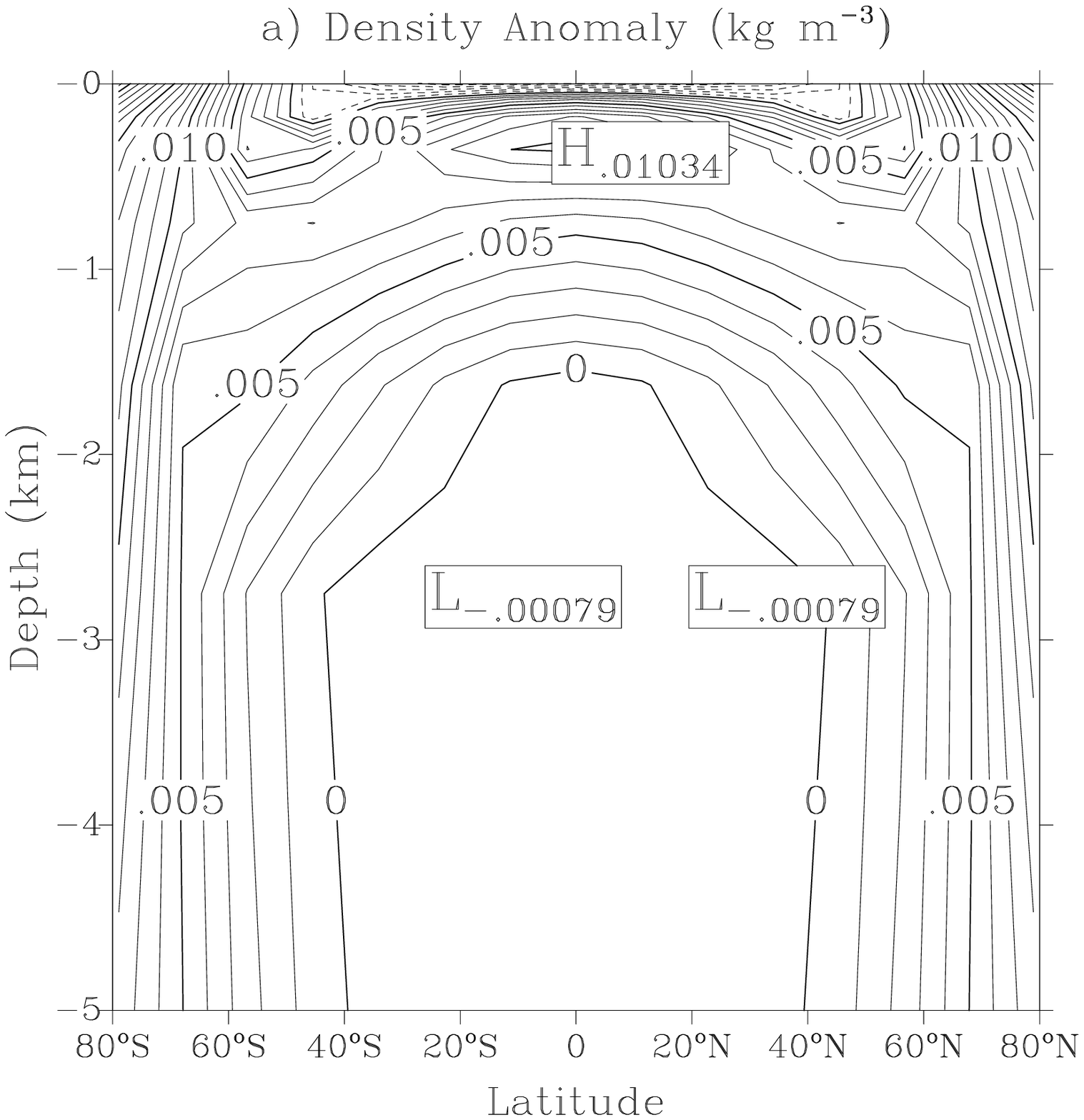}}
\put(200,0){\includegraphics{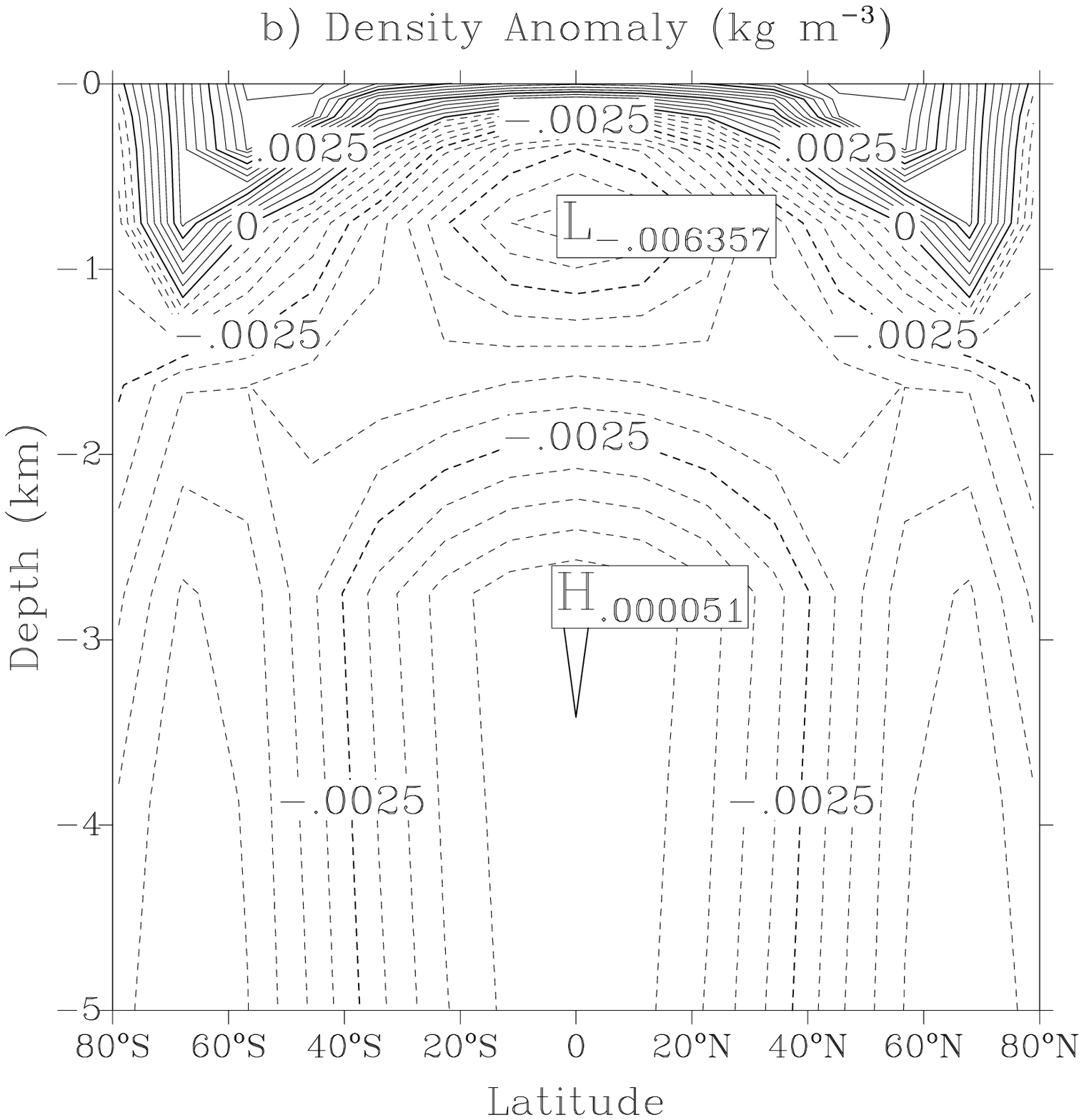}}
\end{picture}
\begin{quote}{\bf
Figure 3}: The real, (a), and imaginary, (b), parts of the
density perturbation for the growing oscillatory eigenfunction for
the two-cell circulation in the (2,1) case. In this case, and in all
other oscillatory cases, the mode is symmetric in temperature,
salinity and density and hence anti-symmetric in the streamfunction
perturbation.
\end{quote}\end{figure}\renewcommand{\baselinestretch}{1.5}\normalsize

\renewcommand{\baselinestretch}{1.0}\small
\begin{figure}[h]
\begin{picture}(400,500)
\put(-72,200){\includegraphics{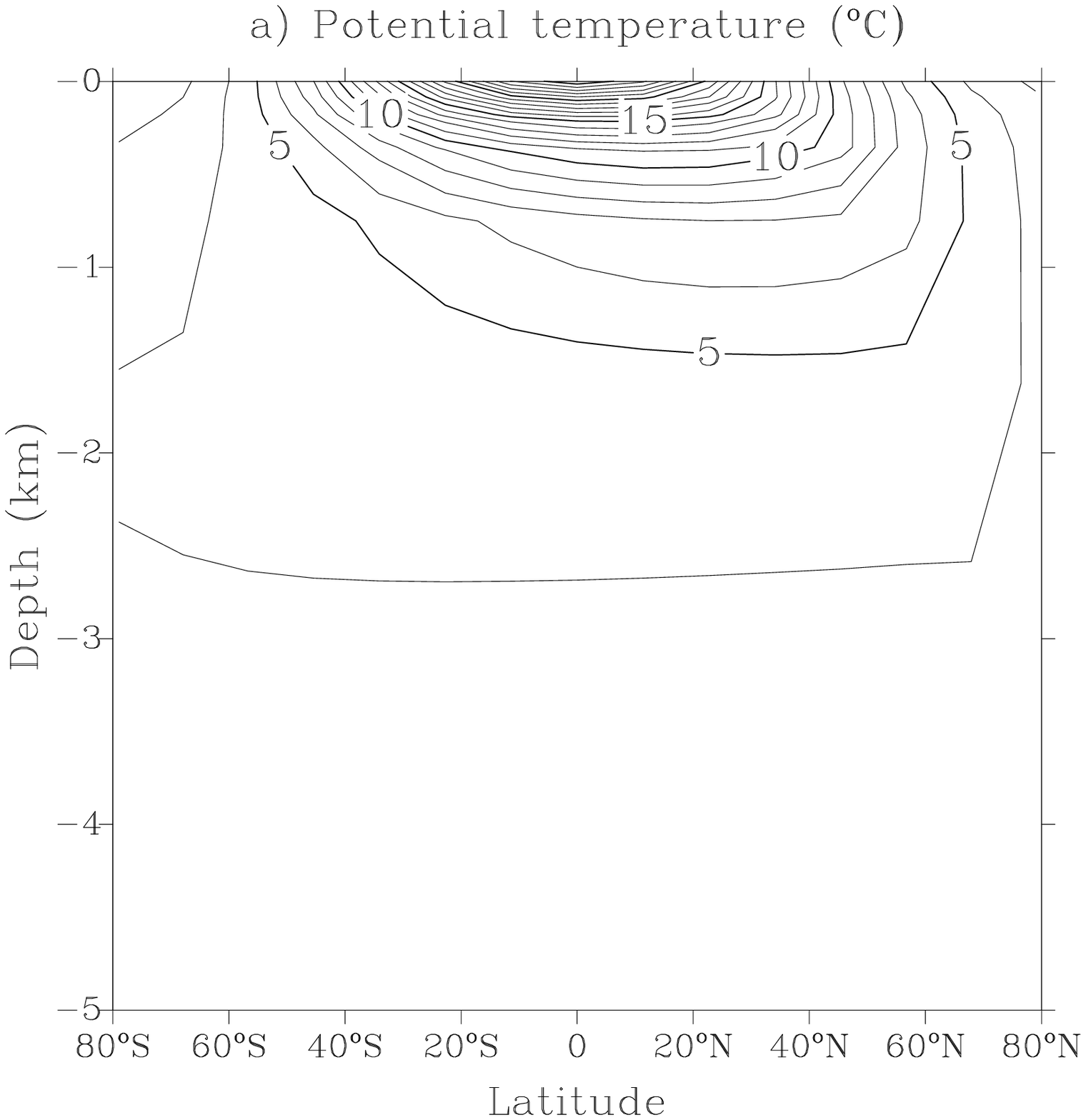}}
\put(200,200){\includegraphics{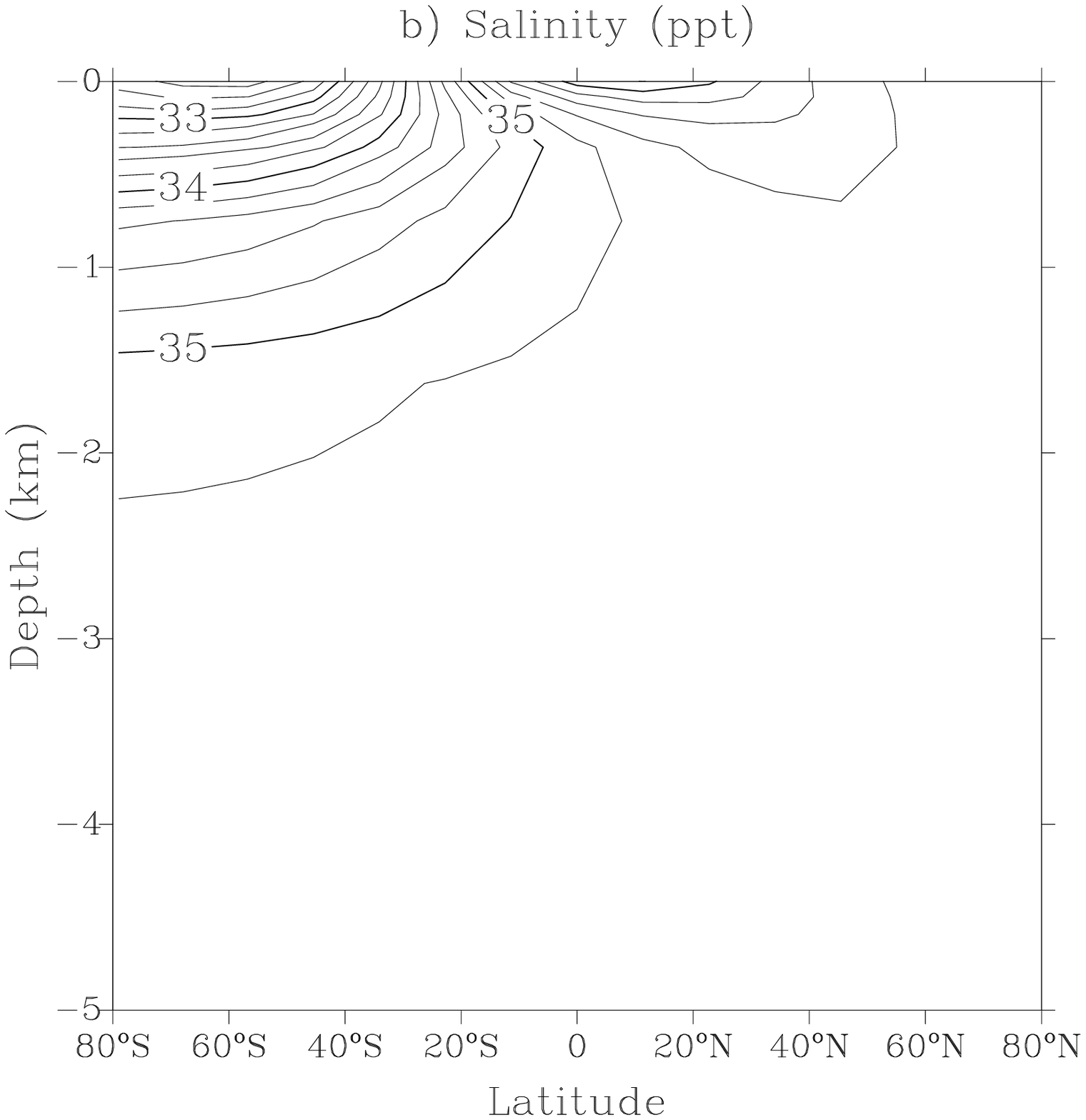}}
\put(-72,-50){\includegraphics{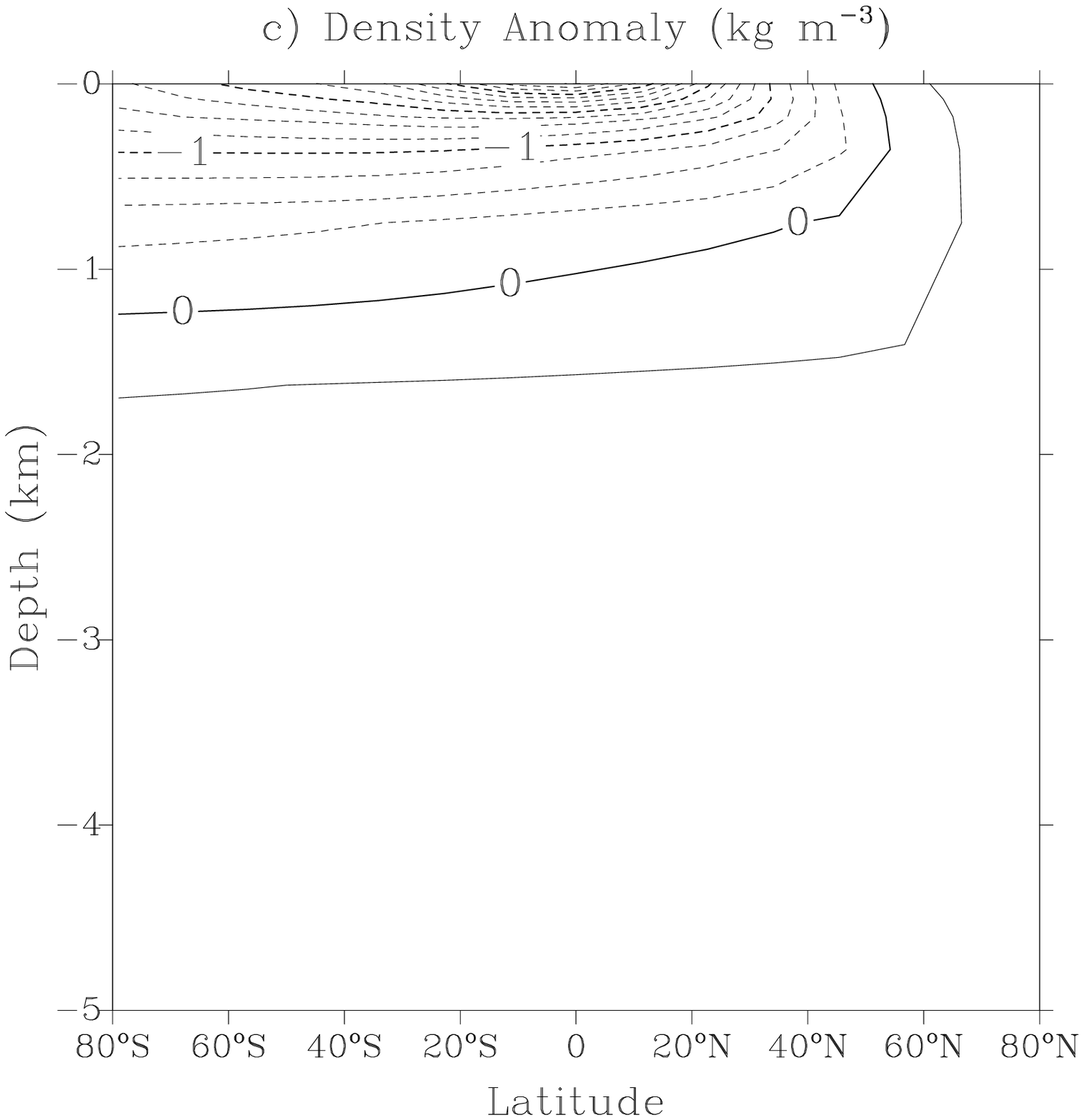}}
\put(200,-50){\includegraphics{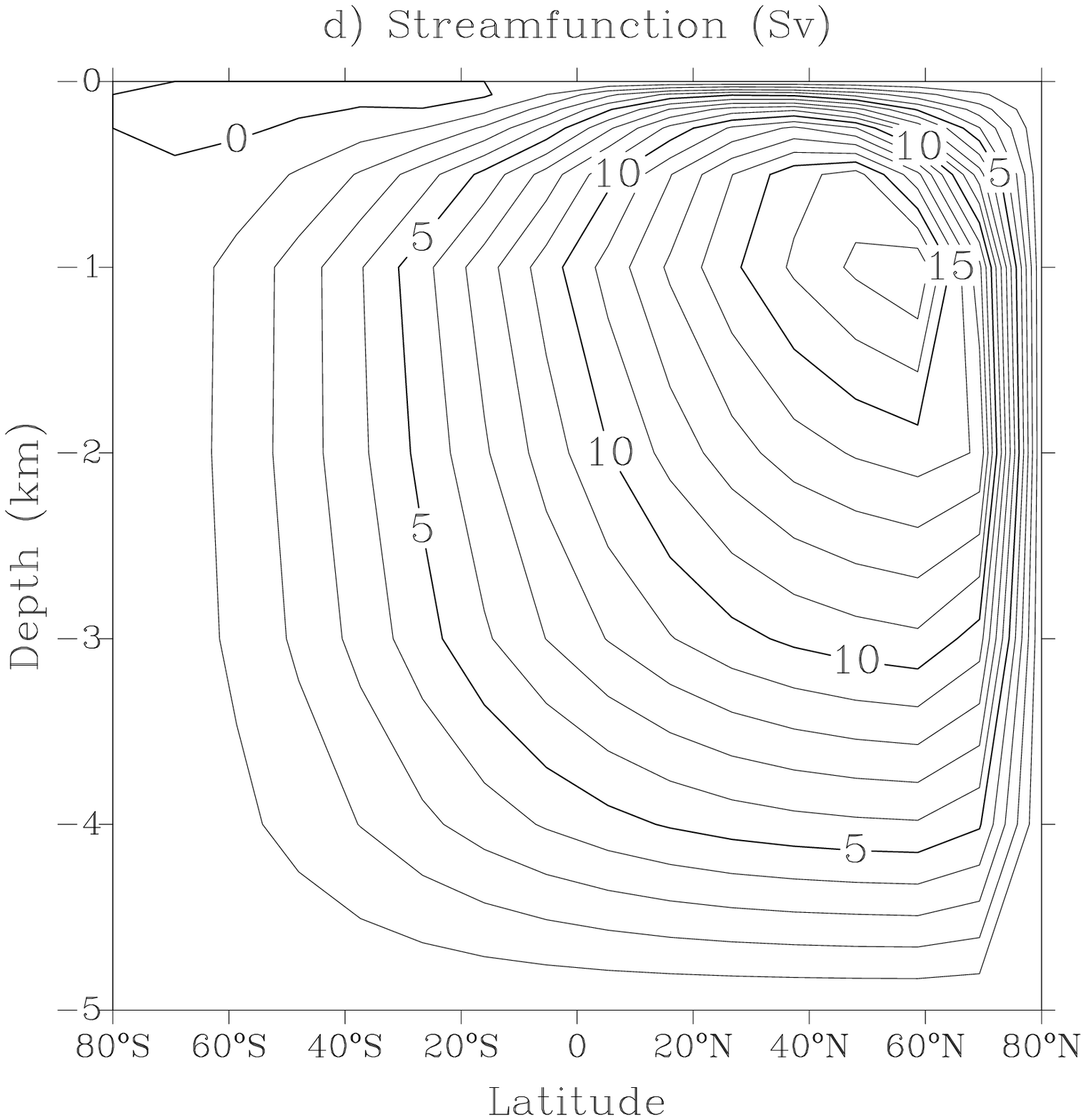}}
\end{picture}
\begin{quote}{\bf
Figure 4}: The (a) Temperature, (b) salinity, (c) density anomaly and (d)
the streamfunction for the northern sinking one-cell circulation
in the (2,1) canonical diffusivities case under mixed boundary conditions.
\end{quote}\end{figure}\renewcommand{\baselinestretch}{1.5}\normalsize

\renewcommand{\baselinestretch}{1.0}\small
\begin{figure}[h]
\begin{picture}(400,500)
\put(-72,200){\includegraphics{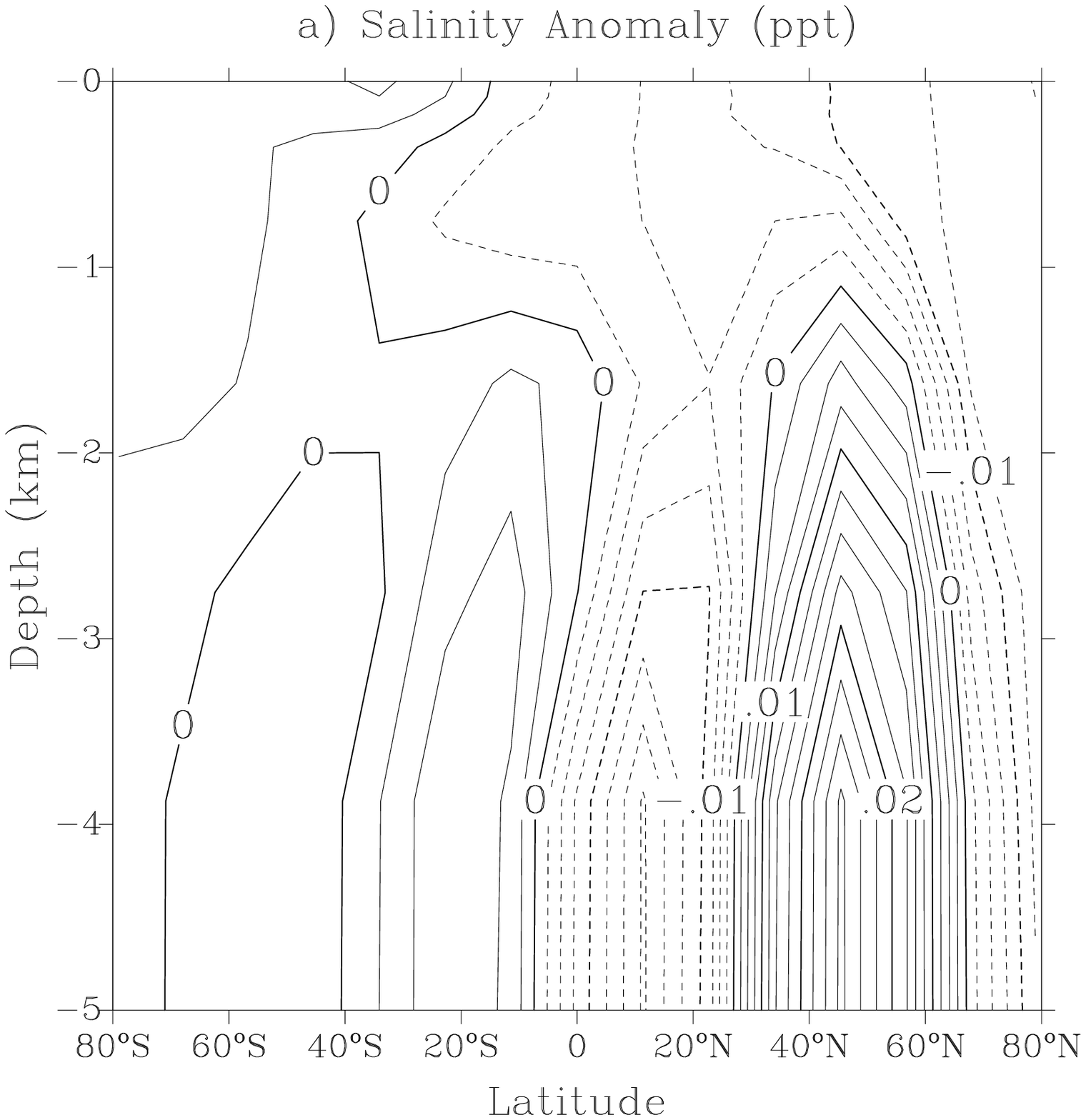}}
\put(200,200){\includegraphics{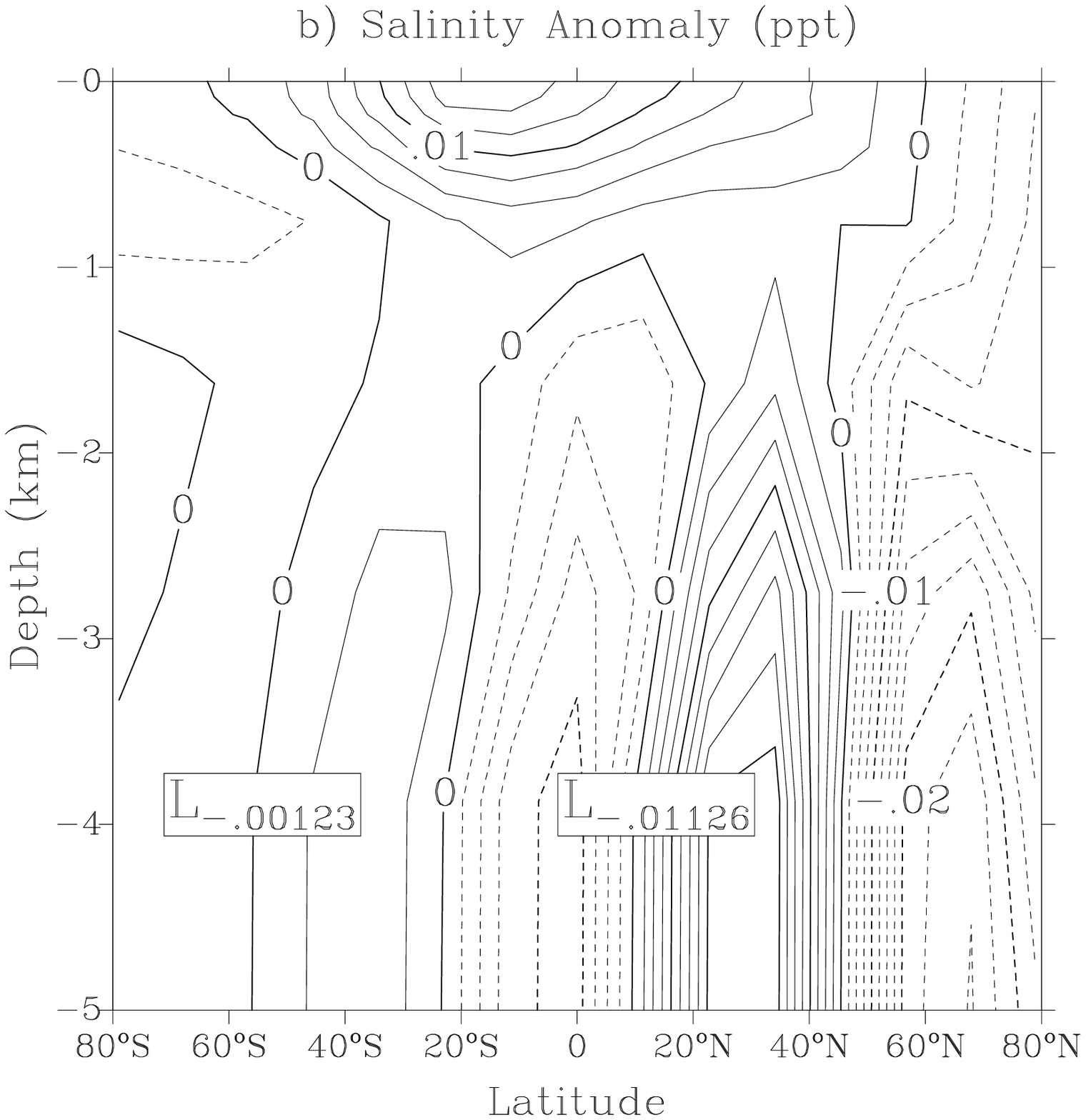}}
\put(200,-50){\includegraphics{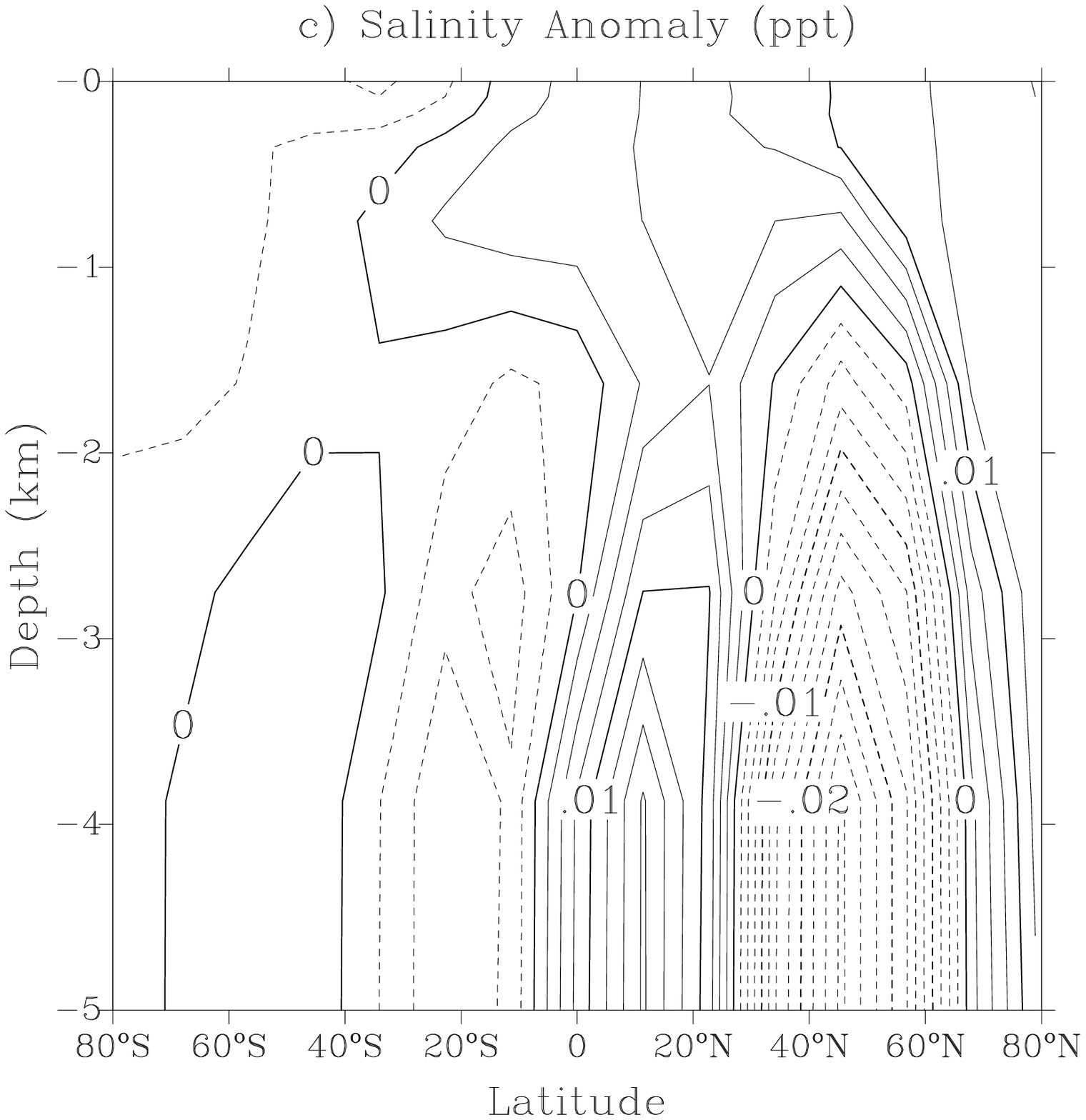}}
\end{picture}
\begin{quote}{\bf
Figure 5}: Snapshots at a) $t=t_0$, b) $t=t_0+\pi/2\omega$ and
c) $t=t_0+\pi/\omega$,  of the salinity anomaly field over a half
period ($\approx$ 150 yrs) of the sub-critical oscillation for the
case (4,1). The following half period is simply this sequence
multiplied by -1. It is clear that the salinity anomalies develop at
the surface and  are then advected clockwise by the main one-cell
circulation.
\end{quote}\end{figure}\renewcommand{\baselinestretch}{1.5}\normalsize

\fig{400,300}{48,0}{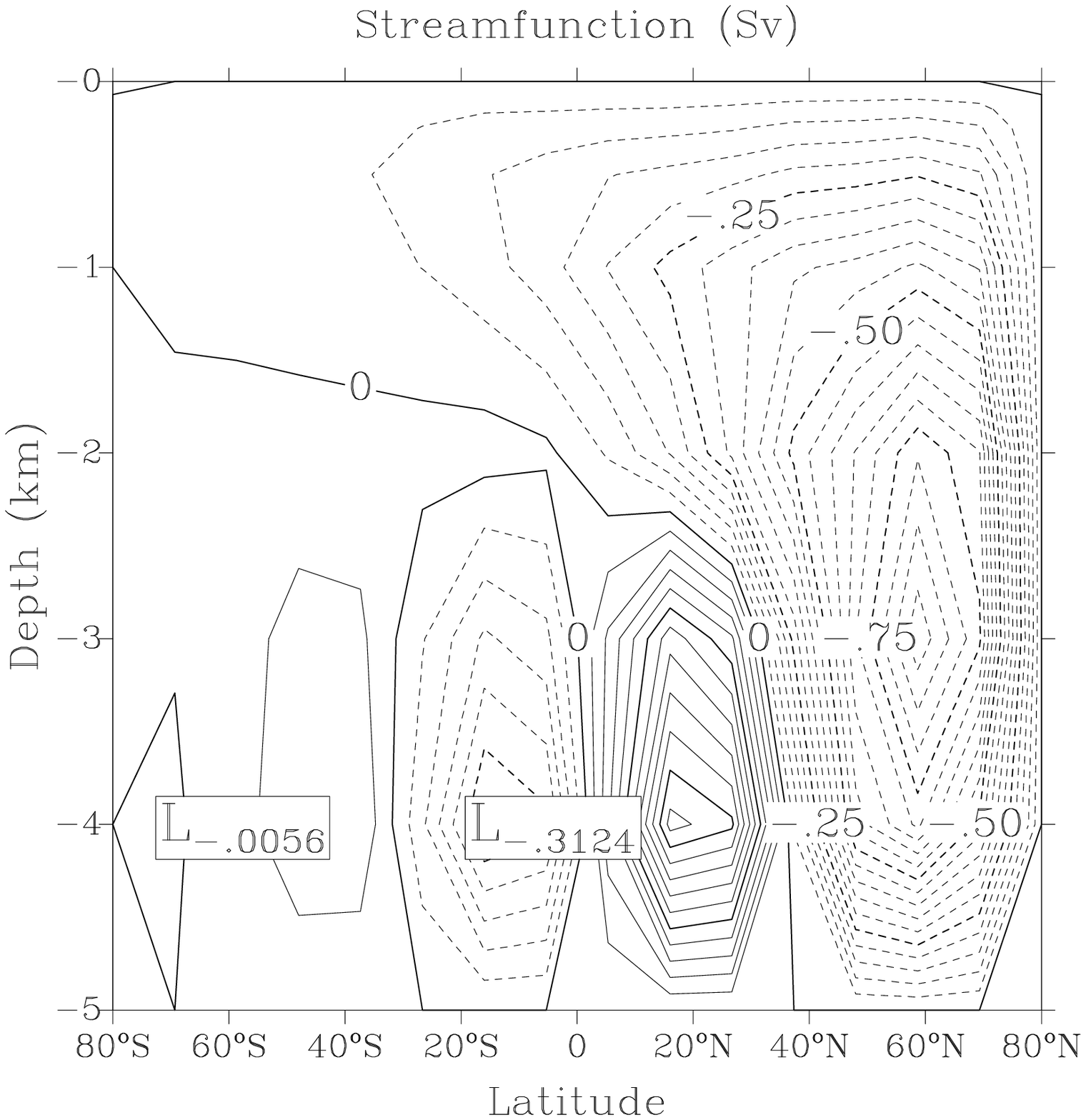}{50.0}{The anomaly in the streamfunction field
for the case
(4,1) due to the initial salinity perturbation show in Fig. 5a. Most
of the perturbation is in the deep ocean where the basic state
temperature and salinity fields are almost constant and hence has only
a minor r\^ole in the propagation of the oscillation.}

\fig{400,300}{48,0}{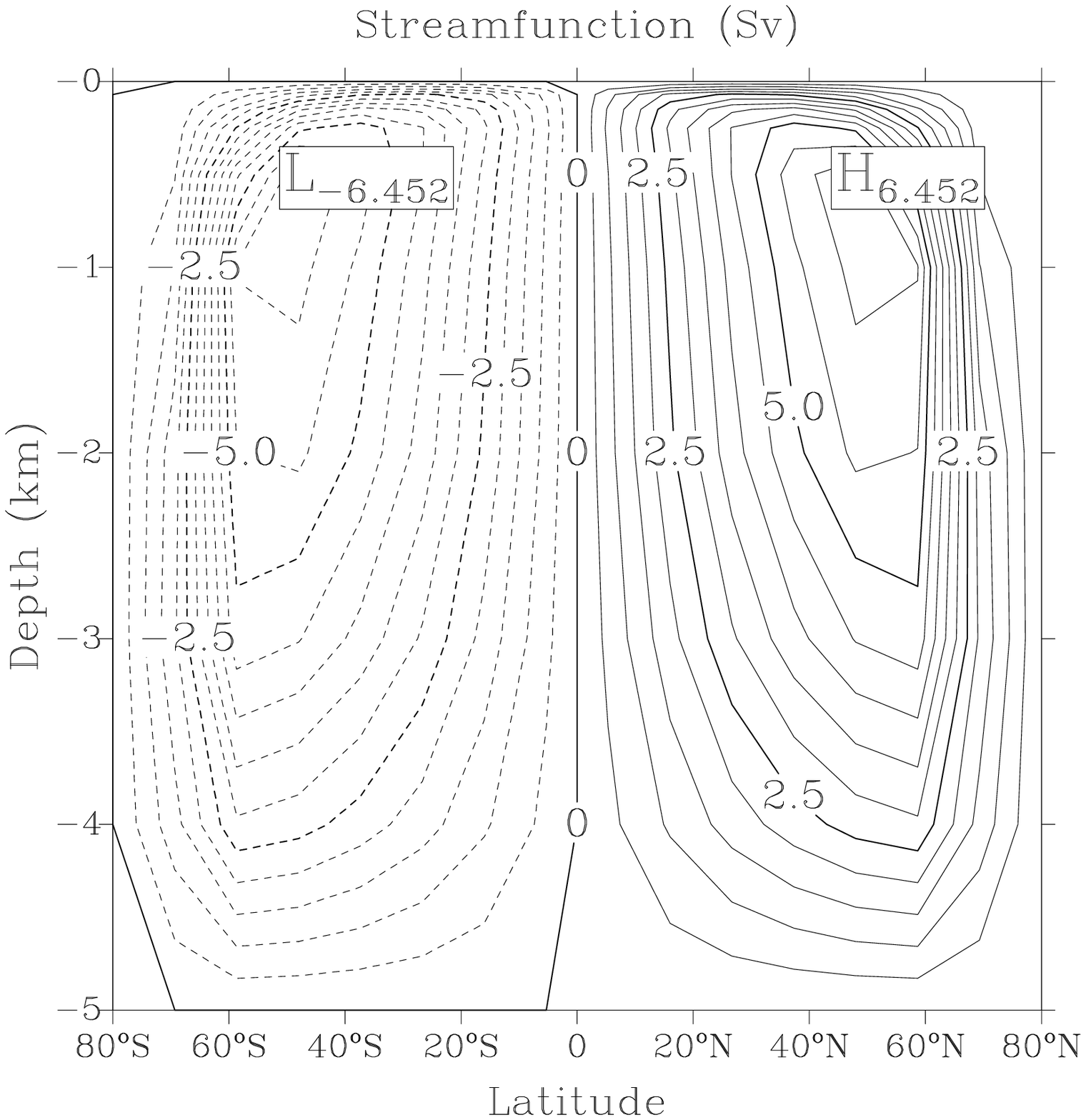}{50.0}{The streamfunction at steady state
under restoring
boundary conditions for the case where $\tau_S=600$ days and the
$\tau_T=50$ days. The basic state has small pools of neutrally stable water
at the poles and most convection and advection occurs in the
neighbouring boxes. The anomalously fast growing perturbations to this
basic state are concentrated solely in the surface polar grid boxes
and are due to the large gradients in the convection
scheme at the point of near-neutral stability.}

\fig{400,354}{50,50}{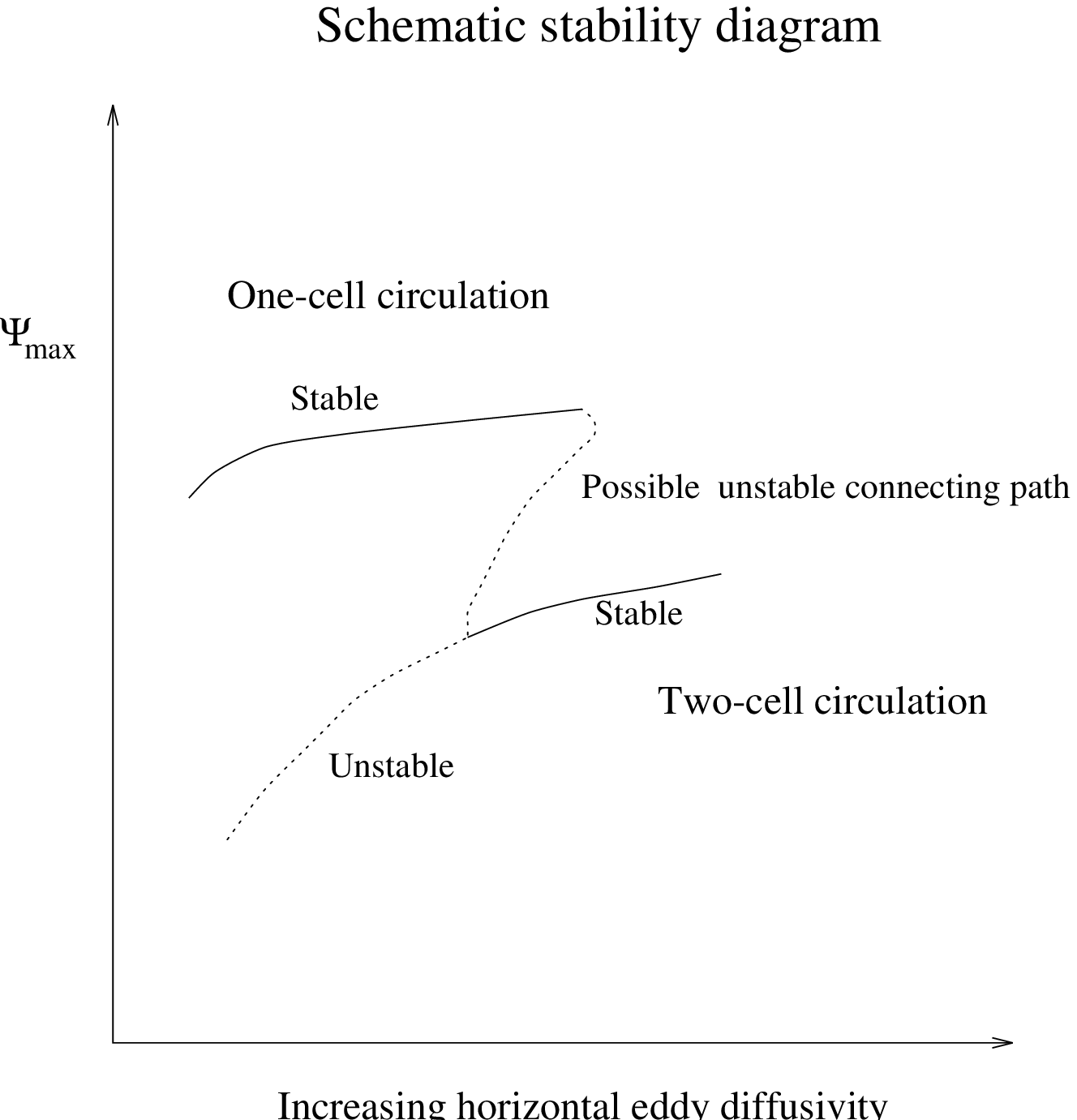}{80.0}{A schematic diagram showing the
bifurcation structure of the model as the eddy diffusivities change.
The horizontal diffusivity is increased while keeping the $K_v$ and
$K_h$ at a constant ratio. The two steady states represented are the
northern-sinking one-cell circulation (always stable if present) and
the symmetric two-cell circulation. The two-cell pattern becomes
stable as $K_h$ increases and we conjecture that there is a pitchfork
bifurcation at this point.}
\end{document}